\documentclass{aa}  

\usepackage{graphicx}
\usepackage{subcaption}
\usepackage{txfonts}
\usepackage{multirow}
\usepackage[dvipsnames]{xcolor}
\begin{document}

   \title{A comparative analysis of pulse time-of-arrival creation methods}

   \author{J.~Wang
          \inst{1}\fnmsep\thanks{jun.wang.ucas@gmail.com}
          \and
          G.~M.~Shaifullah
          \inst{2}
          \and
          J.~P.~W.~Verbiest
          \inst{1,3}
          \and
          C.~Tiburzi
          \inst{3,4}
          \and
          D. J. Champion
          \inst{3}
          \and
          I. Cognard
          \inst{5}
          \and
          M. Gaikwad
          \inst{3}
          \and
          E. Graikou
          \inst{3}
          \and
          L. Guillemot
          \inst{5}
          \and
          H. Hu 
          \inst{3}
          \and
          R. Karuppusamy 
          \inst{3}
          \and
          Michael J. Keith 
          \inst{6}
          \and
          Michael Kramer 
          \inst{3,6}
          \and
          Y.~Liu
          \inst{1}
          \and
          A.~G.~Lyne
          \inst{6}
          \and
          M. B. Mickaliger 
          \inst{6}
          \and
          B. W. Stappers
          \inst{6}
          \and
          G. Theureau
          \inst{5,7}
   }

   \institute{Fakulat\"at f\"ur Physik, Universit\"at Bielefeld,
     Postfach 100131, 33501 Bielefeld, Germany
     \and
     Dipartimento di Fisica `G. Occhialini', Universit\'{a} degli Studi di Milano-Bicocca, Piazza della Scienza 3, 20126 Milano, Italy; INFN, Sezione di Milano-Bicocca, Piazza della Scienza 3, 20126 Milano, Italy
     \and
     Max-Planck-Institut f\" ur Radioastronomie,
     Auf dem H\" ugel 69,
     53121 Bonn, Germany
     \and
     ASTRON, the Netherlands Institute for Radio Astronomy,
     Postbus 2, 7990 AA Dwingeloo, The Netherlands
     \and
     Laboratoire de Physique et Chimie de l'Environnement et de l'Espace LPC2E CNRS-Universit{\'e} d'Orl{\'e}ans, 45071, Orl{\'e}ans, France; 	
	 Station de radioastronomie de Nan{\c c}ay, Observatoire de Paris, PSL Research University, CNRS/INSU 18330 Nan{\c c}ay, France 
     \and
     Jodrell Bank Centre for Astrophysics, University of Manchester,
     Manchester, M13 9PL, United Kingdom 
     \and
     Laboratoire Univers et Th[\'e]ories LUTh, Observatoire de Paris, PSL Research University, CNRS/INSU, Universit{\'e} Paris Diderot, 5 place Jules Janssen, 92190 Meudon, France 
   }

   \date{Received MMMMM DD, YYYY; accepted MMMMM DD, YYYY}

   
   \abstract
   {Extracting precise pulse times of arrival (TOAs) and their
     uncertainties is the first and most fundamental step in
     high-precision pulsar timing. In the classical method, TOAs are derived from total intensity pulse profiles of pulsars via cross-correlation with an idealised `1D' template of that profile. While a number of results have been presented in the literature relying on the ever increasing sensitivity of such pulsar timing experiments, there is no consensus on the most reliable methods for TOA creation and, more importantly, the associated TOA uncertainties for each scheme. 
     }
   {In this article, we present a comprehensive comparison of TOA
     determination practices, focusing on the creation of
     timing templates, TOA determination methods and the most useful TOA bandwidth.
     The aim is both to present a possible approach towards TOA optimisation as well as the (partial) identification of an optimal TOA-creation scheme and the demonstration of optimisation differences between pulsars and data sets.}
   {We compare the values of data-derived template profiles as compared
     to analytic profiles and evaluate the three most commonly used
     template-matching methods. Finally, we study the relation
     between timing precision and TOA bandwidth to identify any
     potential breaks in that relationship. As a practical
     demonstration, we apply our selected methods to European Pulsar
     Timing Array data on the three test pulsars PSRs\ J0218+4232,
     J1713+0747 and J2145$-$0750.}
   {Our demonstration shows that data-derived and
   smoothed templates are typically preferred to some more commonly
   applied alternatives; the Fourier domain with Markov-chain Monte
   Carlo (FDM) template-matching method is generally superior to or
   competitive with other methods; and while the optimal TOA
   bandwidth is strongly dependent on pulsar brightness, telescope
   sensitivity and scintillation properties, some significant frequency
   averaging seems required for the data we investigated.}
   {}
   

   \maketitle
%

\section{Introduction}\label{sec:intro}

Millisecond pulsars \citep[MSPs, first discovered
  by][]{bkh+82} are neutron stars which have been spun
up or "recycled" to rotation periods shorter than $\sim$30\,ms through
accretion of matter from a binary companion star
\citep{acrs82,bv91}. Their extremely stable
rotation periods make MSPs ideal laboratories to test a diversity of
extreme astrophysical phenomena that cannot be accessed on
Earth. Their applications include investigating turbulence and structures of the interstellar
medium \citep{kcs+13,leg+18,dvt+19}, testing relativistic gravity
\citep{tw82,agh+18,vcf+20}, building a pulsar-based time standard
\citep{hcm+12,hgc+20}, constraining masses in the Solar System \citep{chm+10,cgl+18},
measuring the magnetic field structure in the Galaxy \citep{hmvd18,gmd+18}
and detecting gravitational radiation \citep[e.g.\ ][]{srl+15,lsc+16,abb+20}.

Detecting nanohertz gravitational waves (GWs) using pulsar timing is
one of the main foci in pulsar-timing research at present \citep[see
  the recent reviews by][]{tib18,btc+19}. It had been predicted
\citep{jhlm05} that a detection can be achieved by monthly
observations of 20 to 40 MSPs over five to ten years if these sources
are characterised by white timing residuals with a root-mean-square
(RMS) of $\sim100$\,ns.  In order to achieve this ambition, three
major Pulsar Timing Arrays \citep[PTAs, see][]{fb90,rom89} were
constructed: the Parkes Pulsar Timing Array \citep[PPTA,][]{mhb+13},
the European Pulsar Timing Array \citep[EPTA,][]{jsk+08, dcl+16} and the
North-American Nanohertz Observatory for Gravitational waves
\citep[NANOGrav,][]{dfg+13}. These three were subsequently combined in
the International Pulsar Timing Array \citep[IPTA,][]{vlh+16,pdd+19}.  Even
though substantial progress has been achieved in improving timing
precision and sensitivity in subsequent data releases, both at the
regional and global level, a detection has so far not been achieved,
primarily due to two limiting factors. Firstly, even the most
up-to-date and highest-quality data sets \citep[like, e.g.\ the
  NANOGrav data set from][]{aab+21a, aab+21b} have too few pulsars with timing
precision below or at the 100-ns level. Secondly, there are a number of corrupting effects, whose impact on timing precision requires careful consideration \citep{sc12, vs18}. One example of
recent progress in that regard, is the influence of the Solar-System
Ephemerides, which was known to affect GW experiments \citep{thk+16},
but it was not properly taken into account until more recently
\citep[e.g.\ ][]{abb+18a, vts+20}. Beyond this, several fundamental aspects of
pulsar-timing analyses were identified as potential targets for
optimisation and harmonisation efforts during the first global PTA
data combination \citep{vlh+16}. Three particular recommendations from
that work will be described below and investigated in detail
throughout this paper.

Firstly, all standard pulsar-timing analyses are based on matching a
so-called template or ``standard'' profile to the observations.  There
are, however, several different ways in which such template profiles
can be constructed. As outlined by \citet{lk05}, traditionally
template profiles are constructed through addition of a large number
of observations, thereby resulting in a pulse profile with a far
higher signal-to-noise ratio (S/N) than any given observation. Such
templates have the advantage that they by definition fully resemble
the actual pulse shape, but the noise of the original observations is
also contained in the template and hence may cause inaccurate
correlations commonly referred to as ``self-standarding''
\citep{hbo05a}. Three strategies have been devised to mitigate this issue: firstly, the template profile can be smoothed;
secondly the template can be restricted to contain only the single
brightest observation (which is then removed from the subsequent
timing analysis and only used as template profile); and thirdly the
template can be modelled with noise-free analytic functions
\citep{lk05}. As can be seen in, e.g.\ \citet{dcl+16,dfg+13,krh+20}
several of the above-mentioned approaches are presently in use, but no
comprehensive comparison of the various template-creation options has
been published.

Secondly, after a template profile has been created, the time-of-arrival (TOA) 
of a given observation is derived from the
cross-correlation between the observation and the
template. Traditionally this has been done in the Fourier domain, by
fitting the phase-gradient of the cross-power spectrum of the template
and the observation 
\citep{tay92}, commonly referred to as the “Phase Gradient Scheme (PGS)” method. However, the
standard implementation of this approach has often been found to
report underestimated uncertainties.  One suggested solution
determines the TOAs in the same way, but derives the TOA uncertainties
from a Monte Carlo analysis (referred to as the ``FDM''
method\footnote{Whilst technically it is accurate to refer to cross-correlation algorithms uniquely in terms of their TOA determination, throughout this paper we refer to cross-correlation \emph{methods} in terms of both their TOA and TOA uncertainty determination. This is in line with how these algorithms are used in practice; and recognises the practical constraint that measurements and their uncertainties must be inextricably linked. Mathematically speaking, therefore the pure TOA-determination algorithms of FDM and PGS are identical, but the cross-correlation methods (which we consider to include the measurement uncertainty as well) will be considered different.}). This latter solution has been suggested \citep{vlh+16} as
being superior. Furthermore, \citet{hbo05a} proposed a
Gaussian-interpolation (``GIS'') cross-correlation algorithm (CCA) in the
time-domain, which is supposedly better suited to low-S/N data
sets. While recent pulsar-timing efforts have increasingly adopted the FDM method, usage of the PGS method is still widespread \citep{dcl+16}, in some cases \citep[e.g.\ ][]{abb+15} with lower bounds imposed on the pulse S/Ns in order to avoid  underestimation of TOA
uncertainties. Here too a comprehensive, data-based (as opposed to simulation-based) comparative evaluation of the
possible alternatives has so far not been undertaken.

%
%
%
Lastly, potential corruptions in the derived TOAs can arise from the
usage of data-recording systems with large (${\gtrapprox}0.3$)
fractional bandwidths \citep{vs18}. This is particularly the case if
the pulse-profile changes shape across the band \emph{and} the
scintillation bandwidth is of the same order as (or slightly smaller
than) the observing bandwidth. Both of these conditions appear to be
likely if the fractional bandwidth is large \citep{dhm+15,lmj+16},
causing the frequency-averaged profile shape to vary from one
observation to the next. If the TOAs are derived from
frequency-averaged observations and template, then the aforementioned
variations will be reflected in the TOAs and in the TOA
uncertainties, worsening the overall timing precision. One solution to
this problem is to carry out two-dimensional template
matching where both observation and template have frequency resolution
and where in addition to the phase offset (or TOA) a frequency drift
(or dispersion measure) is determined \citep{ldc+14,pdr14}. An
alternative solution is to derive multiple TOAs across the bandwidth
of the observation \citep{dfg+13,abb+15}, either against a
frequency-averaged template or through matching each channel of the
observation against the corresponding channel in the template. In all
of this, it is an unsolved question which TOA bandwidth would be
optimal: too narrow a bandwidth would leave little signal and create
TOAs that have high levels of radiometer noise, too wide a
bandwidth and the scintillation effects risk becoming significant -- the
most useful TOA bandwidth is likely strongly pulsar dependent.

To date, PTAs either fully
frequency averaged their TOAs \citep{dcl+16}, used timing-model
extensions to correct for profile-shape variations \citep{abb+15} or
have carried out a mixed approach where both frequency-averaged and
frequency-resolved TOAs have been derived \citep{krh+20}.

The three detailed aspects of TOA creation outlined above are
investigated in this paper. Specifically in Section~\ref{sec:obs} we introduce the data and pulsars we use to
test and compare the aforementioned aspects of timing on these data; in Section~\ref{sec:proc} we describe the various analysis methods that were compared: template creation is
discussed in Section~\ref{ssec:std}, template matching methods in
Section~\ref{ssec:CCA} and most useful TOA bandwidth determination in
Section~\ref{ssec:TOABW}. Before carrying out the actual analysis on the real data, we run some simulations to evaluate the reliability of the TOA uncertainties derived by the various TOA determination methods. These simulations are presented in Section~\ref{sec:simulation}. We discuss our findings on the three test
pulsars in Section~\ref{sec:results} and summarise of our
findings in Section~\ref{sec:conc}.

\section{Observations}\label{sec:obs}

Our analysis is based on data from the newest generation of data
recorders at four of the radio telescopes that constitute the EPTA:
the Effelsberg radio telescope (designated as ``EFF''), the
Jodrell Bank Lovell radio telescope (JBO), the Nan\c cay decimetric
radio telescope (NRT) and the Westerbork synthesis radio telescope
(WSRT). Details on the observing systems used at each of these
observatories are given below. The pulsars used for this initial
investigation are PSRs\ J0218+4232, J1713+0747 and J2145$-$0750. These
three pulsars were chosen because they represent a variety of
characteristics. Specifically, PSR~J1713+0747 is very bright, with a
small duty cycle, and is one of the most precisely timed MSPs in PTAs.
PSR~J0218+4232 has a relatively high dispersion measure (DM), is fainter and consequently not as well timed. PSR~J2145$-$0750 is in between the other two in terms of brightness and timing precision, but has a significantly
longer spin period and appears to have more significant stochastic
wide-band impulse-modulated self-noise \citep[or SWIMS,][commonly also known as pulse-phase jitter]{ovh+11}.

Some basic properties for these three
pulsars are given in Table~\ref{tab:psrs}; the observatory-specific
details of the observations used are given in
Table~\ref{table:pulsae_parameter} and the number and date ranges of
the observations are given in Table~\ref{tab:obs}.

In addition to the telescope-specific radio-frequency interference (RFI) mitigation strategies
described below, all data were visually inspected through the software
suite \textsc{psrchive} \citep{hvm04,vdo12} before TOAs were derived
as described in Section~\ref{sec:proc}. Specifically, since the data 
near the edge of each sub-band are often corrupted by aliasing and spectral 
leakage problems, these channels are excised with the \textsc{paz} program.

\begin{table*}
  \caption{Basic parameters for the pulsars in our sample. Parameters
    are merely indicative and have been rounded. As an indication of
    the timing precision, the weighted RMS residual is quoted for the
    IPTA combination that provided the best timing precision for the
    given pulsar, rounded to one significant digit. ECORR refers to
    the mean ``frequency-correlated EQUAD parameter'' \citep{abb+14} and
    is a measure for the strength of SWIMS present in a given pulsar.}
  \centering
  \label{tab:psrs}
  \begin{tabular}{@{}lrrrl@{}}
    \hline
    Parameter                 & J0218+4232 & J1713+0747 & J2145$-$0750 & References \\
    \hline
    Period (ms)               & 2.3        & 4.6        & 16.1         & \small{\citet{nbf+95,fwc93,bhl+94}}\\
    DM (cm/pc$^3$)            & 61         & 16         & 9            & \small{\citet{nbf+95,fwc93,bhl+94}}\\
    Orbital Period (days)     & 2.0        & 67.8       & 6.8          & \small{\citet{dcl+16,abb+18}}\\
    Flux density S1400 (mJy)  & 0.9        & 9.1        & 10.3         & \small{\citet{kxl+98,dhm+15}}\\
    Pulse Width W50 (\%)      & 43         & 2.4        &  2.1         & \small{\citet{stc99,mhb+13}}\\
    IPTA Timing RMS ($\mu$s)  & 7          & 0.2        & 1            & \small{\citet{vlh+16,pdd+19}}\\
    IPTA ECORR                & --         & 0.16       & 1.3
    & \small{\citet{pdd+19}}\\
    \hline
  \end{tabular}
\end{table*}

\begin{table}
  \caption{Description of the observational set-up for the data used
    in this paper. Given are the telescope identifier (see text), the
    name of the data recording device (referred to as ``backend''),
    the centre frequency $f_{\text{c}}$ of the observations, the
    observing bandwidth $BW$ (not considering RFI or band-edge
    removal), the number of frequency channels $N_{\text{chan}}$ and the
    maximum number of phase bins across the profile $N_{\text{bin}}$. Note that
    for some pulsars JBO and WSRT only had 256 phase bins, as specified in the text.}  \centering
  \begin{tabular}{@{}llcccc@{}} 
    \hline
    Telescope & Backend & $f_{\text{c}}$  & $BW$  & $N_{\text{chan}}$ & $N_{\text{bin}}$ \\
              &         &       (MHz)     & (MHz) &                   &                   \\
    \hline
    EFF      &  PSRIX  & 1347.5 & 200 & 128         & 1024\\
    JBO      &  Roach  &  1532  & 400 & 400 / 1600 & 2048\\
    NRT      &  NUPPI  &  1484  & 512 & 128         & 2048\\
    WSRT     & PuMa II &  1380  & 160 & 512         & 1024\\
    \hline
    \label{table:pulsae_parameter}
  \end{tabular}
\end{table}

\begin{table}
  \caption{Summary of observations used in this paper. For each
    pulsar-telescope combination the number of observations
    $N_{\text{obs}}$, MJD range and observing time span $T$ is given.}
  \centering
  \label{tab:obs}
  \begin{tabular}{@{}llrcc@{}}
    \hline
    PSR        & Tel ID & $N_{\rm obs}$ & MJD          & $T$ (yrs) \\
    \hline
    J0218+4232 & EFF    &  53           & 55600--57705 & 5.8 \\
               & JBO    & 128           & 55666--58559 & 7.9 \\
               & NRT    & 467           & 55854--58558 & 7.4 \\
               & WSRT   &  65           & 54775--57131 & 6.5 \\
    \\
    J1713+0747 & EFF    &  96           & 55633--58713 & 8.4 \\
               & JBO    & 301           & 55905--58610 & 7.4 \\
               & NRT    & 368           & 55800--58558 & 7.6 \\
               & WSRT   & 112           & 54155--57195 & 8.3 \\
    \\
    J2145$-$0750 & EFF  &  62           & 55633--57733 & 5.7 \\
               & JBO    & 122           & 55880--58495 & 7.2 \\
               & NRT    & 265           & 55803--58555 & 7.5 \\
               & WSRT   &  85           & 54520--57135 & 7.2 \\
    \hline
  \end{tabular}
\end{table}

\subsection{Effelsberg radio telescope}

The Effelsberg radio telescope is a homological Gregorian design with
a 100-m paraboloidal primary reflector and a 6.5-m ellipsoidal
secondary reflector, which makes the Effelsberg radio telescope the
world's second-largest fully movable telescope. It operates at
wavelengths from about 90\,cm to 3.5\,mm (i.e.\ observing frequencies
from 300\,MHz to 90\,GHz).

The data from Effelsberg presented in this research were acquired with
a typical cadence of 3--4 weeks between 2011 and 2019 with an
integration time per source of about 30\,min per observation. The data
recorder (henceforth referred to as ``backend'') used was the PSRIX
pulsar-timing system \citep{lkg+16}, which is based on the Reconfigurable Open Architecture Computing Hardware (ROACH) boards
and carries out coherent dedispersion on central processing units (CPUs) using the \textsc{dspsr}
\citep{vb11} package. The observations were taken with the
\mbox{P-217} and P-200 receivers, centred at 1.3475\,GHz in a
frequency-multiplexing mode where eight 25-MHz-wide subbands are
independently dedispersed and written to disk. This results in
observations with a total of 200\,MHz of bandwidth, 128 frequency
channels, 1024 phase bins, 10-second
integrations and full polarisation information.

These data were combined in time and frequency with the
\textsc{psradd} command of the \textsc{psrchive} package
\citep{hvm04,vdo12}. Subsequently
RFI was removed automatically with 
\textsc{clean.py}, the RFI excision script of the \textsc{COASTGUARD} package
\citep{lkg+16}, using the ``surgical'' algorithm.

\subsection{Lovell radio telescope}

At Jodrell Bank Observatory the 76-m Lovell telescope is used in a
regular pulsar-monitoring programme. The observations used in this work were taken from that programme, have a centre frequency of
1.532\,GHz, a timespan of $\sim$7.5 years, a
typical cadence of three weeks and a typical integration time of
15\,minutes. For PSR~J1713+0747 the cadence was closer to one
week. The backend used for these observations was a ROACH-based
backend very similar to the one used at the Effelsberg observatory and
described by \citet{lkg+16}, and using the \textsc{dspsr} package
\citep{vb11} for real-time coherent de-dispersion. 

The JBO observations have a 
bandwidth of 400\,MHz split over 25 16-MHz-wide subbands, each
channelised into 1 or 0.25\,MHz wide frequency channels. The number of phase
bins varies between 256 and 2048 depending on the pulsar observed. For the data set we used in this work, PSR~J0218+4232 has the minimum 256 bins. 

To remove the RFI, the spectral kurtosis method for real-time RFI
removal \citep{ngl+07} was applied during the \textsc{dspsr}
pre-processing, after which the data were run through some basic
RFI-removal scripts and finally were visually inspected for excluding
the remaining RFI.

\subsection{Nan{\c c}ay decimetric radio telescope}

Built in 1965, the Nan{\c c}ay decimetric radio telescope (NRT) is a
Kraus-type telescope with a collecting area of 6912\,m$^{2}$,
equivalent to a 94-m parabolic dish. Since 2011, NRT observations of
pulsars are done with the ROACH-based NUPPI backend \citep{ctg+13}, 
a version of the Green Bank Ultimate Pulsar Processing 
Instrument \citep{drd+08} designed for Nan\c{c}ay.
With a total bandwidth of 512\,MHz, the NUPPI
backend splits the full bandwidth into 128 channels, each channel with
4\,MHz bandwidth. The observations used here were centred at
1484\,MHz and were coherently dedispersed in real-time on GPU units
that are part of the NUPPI system.

The pulsars were observed with NUPPI since 2011 with a cadence that
varied between five and ten days and with an integration time between
45 and 60 minutes per observation. Various automated RFI excision
schemes were applied in post-processing offline. 

\subsection{Westerbork synthesis radio telescope}

The Westerbork synthesis radio telescope (WSRT), located in the
Netherlands and operated by the Netherlands Institute for Radio
Astronomy (ASTRON), is an east-west interferometric array with 14 25-m
parabolic dishes, which in terms of collecting area is equivalent to a
94-m single telescope.

Pulsar monitoring at the WSRT has been carried out at 350\,MHz,
1.38\,GHz and 2.27\,GHz with the Pulsar Machine II (PuMa-II)
backend \citep{ksv08}. Pulsar observations are recorded in eight 20-MHz-wide
sub-bands, for a total observing bandwidth of 160\,MHz with 512
($=8\times 64$) frequency channels. Coherent dedispersion is performed
offline with the \textsc{dspsr} package.
 
The observations used in this paper were taken at a centre frequency
of 1380\,MHz, had a cadence of roughly one month and a total
integration time of about 30\,minutes. For PSRs\ J0218+4232 and J1713+0747, data with only 256 bins were used in this work.

\section{Data processing techniques}\label{sec:proc}

In this section we describe the different algorithms and methods we
use and compare. Specifically, the template creation methods are
described in Section~\ref{ssec:std}, the template-matching methods
are outlined in Section~\ref{ssec:CCA} 
and the analysis to determine the most useful TOA bandwidth is described
in Section~\ref{ssec:TOABW}. All of these algorithms and methods make
use of the \textsc{psrchive} and \textsc{tempo2} \citep{hem06}
software packages.
 
\subsection{Template creation and comparison}\label{ssec:std}

The ideal template describes the pulse profile shape in perfect
detail, without additive noise. A vital and complicating aspect in
this regard is the fact that the achievable timing precision for a
given profile shape is a strong function of the higher-frequency
components present in the profile shape, i.e.\ of sharp features in
the profile \citep[e.g.\ ][]{van06}. Consequently, depending on the pulsar in
question, noise-mitigation methods like smoothing or modelling of
template profiles may require particular care in order to prevent
negative impacts on the timing potential of a given pulsar.

\begin{figure}
  \centering
  
\begin{subfigure}[b]{0.8\columnwidth}
  \centering
  \includegraphics[width=\columnwidth]{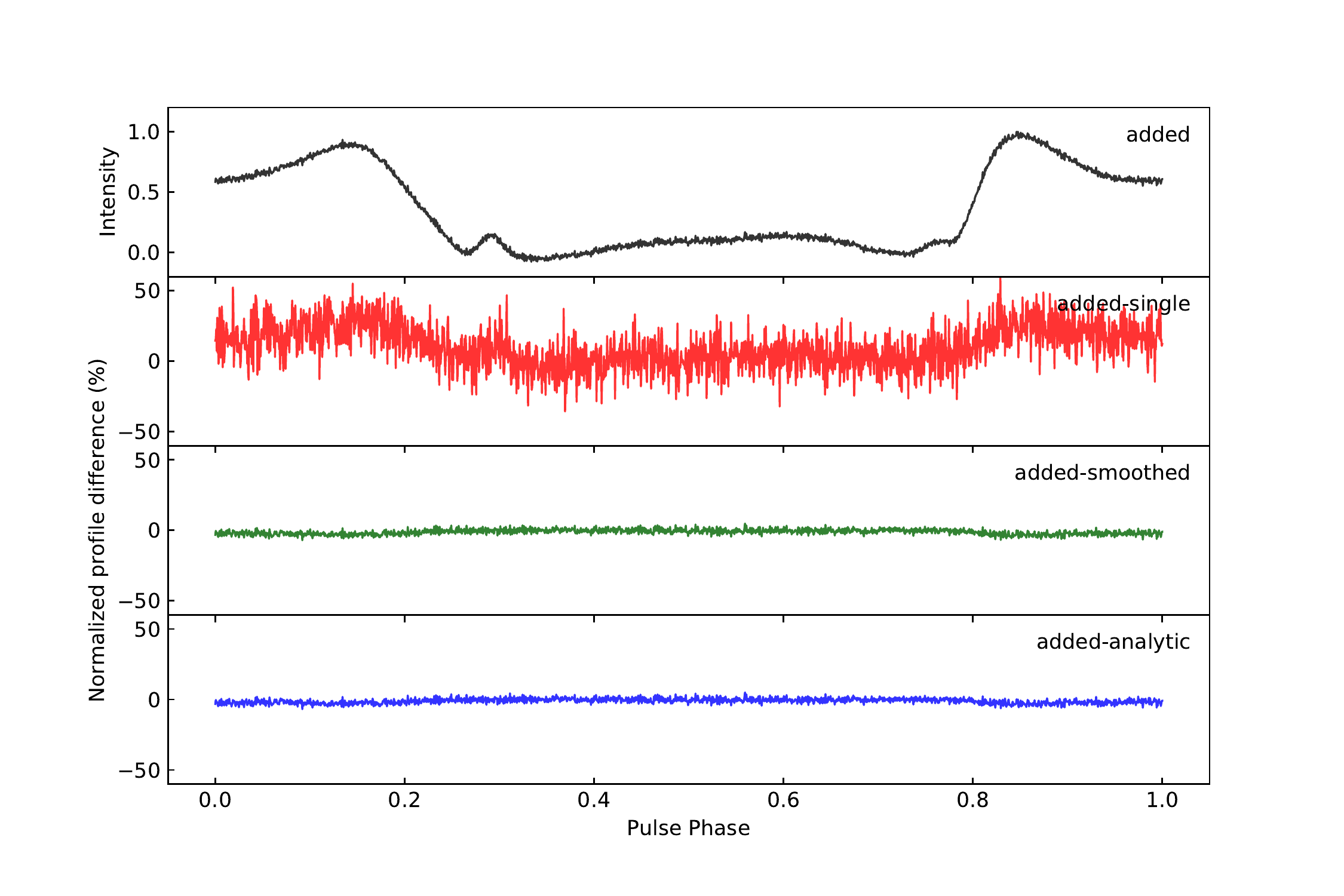}
  \vspace*{-1.5\baselineskip}
  \caption{PSR~J0218+4232.}
  \label{fig:J0218_ncy_profiles}
\end{subfigure}

\begin{subfigure}[b]{0.8\columnwidth}
  \centering
  \includegraphics[width=\columnwidth]{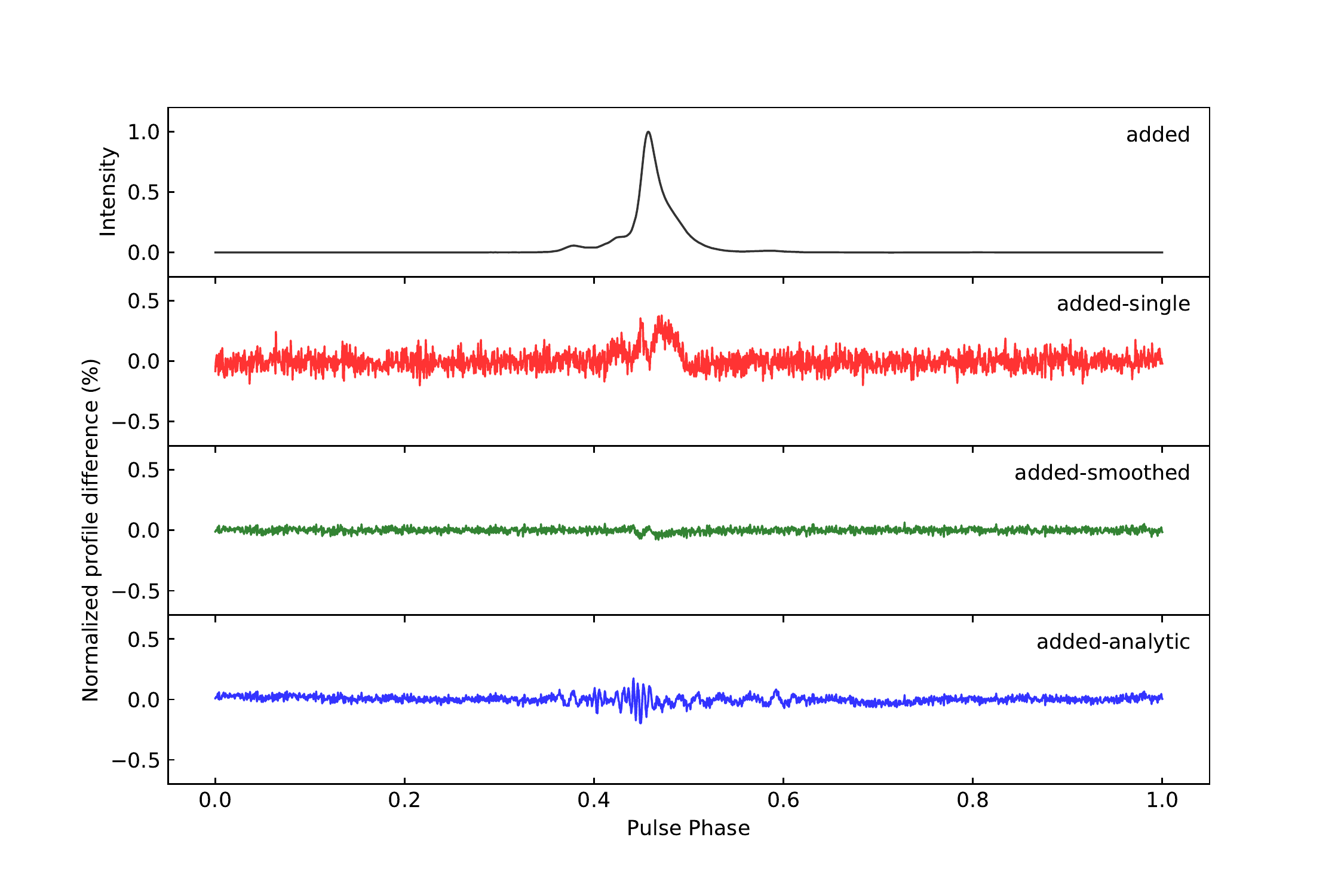}
  \vspace*{-1.5\baselineskip}
  \caption{PSR~J1713+0747.}
  \label{fig:J1713_ncy_profiles}
\end{subfigure}

\begin{subfigure}[b]{0.8\columnwidth}
  \centering
  \includegraphics[width=\columnwidth]{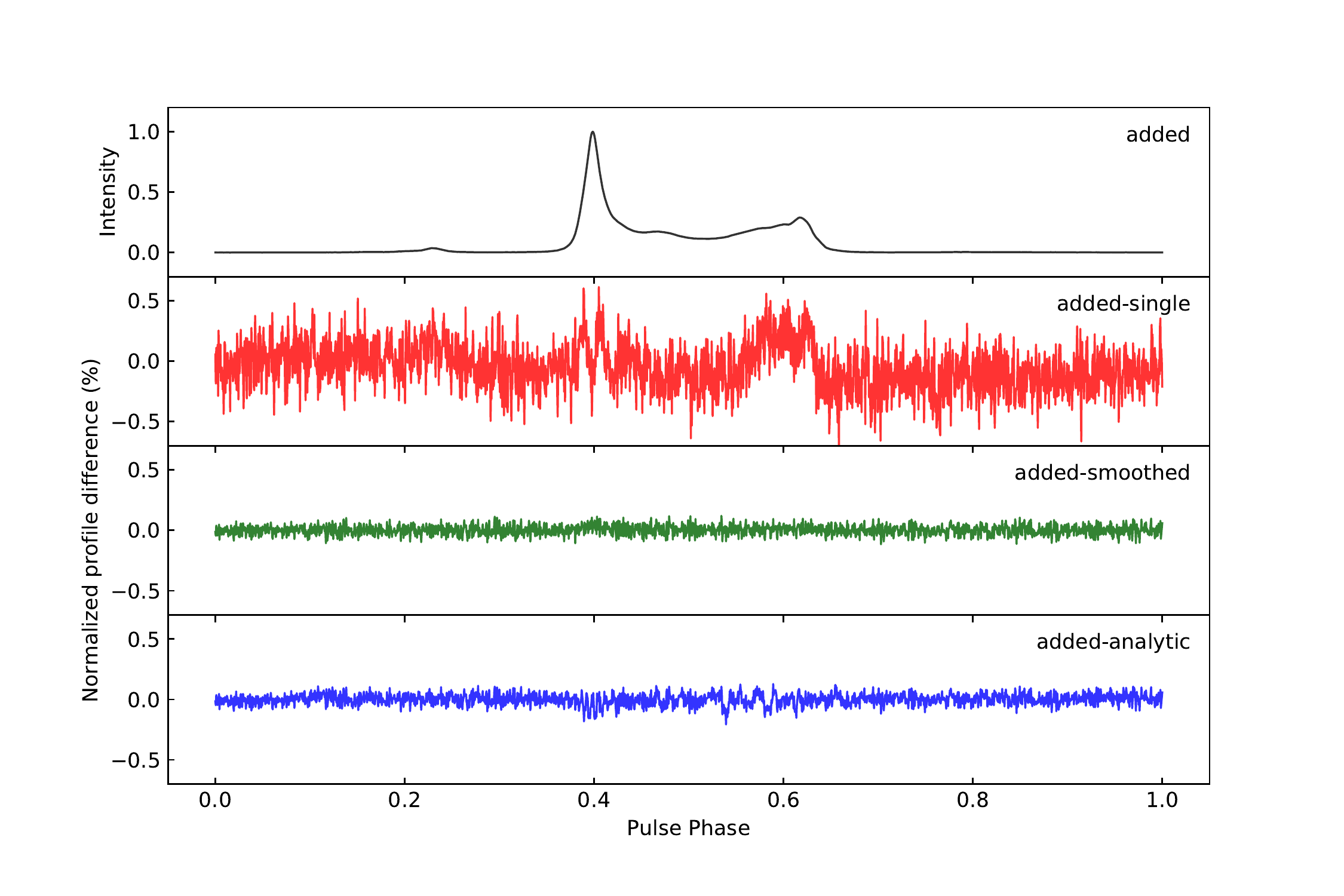}
  \vspace*{-1.5\baselineskip}
  \caption{PSR~J2145$-$0750.}
  \label{fig:J2145_ncy_profiles}
\end{subfigure}

\caption{Peak-normalised added template profile and profile differences between the added and all other templates derived from NRT data. From top to bottom, shown are PSRs\ J0218+4232, J1713+0747, J2145$-$0750.}
\label{fig:ncy_profile}
\end{figure}

Common approaches towards template generation were already discussed
in the introduction, thus here we summarise the four
different methods we compare:
\begin{description}
  \item[\textbf{Single-brightest observation} (henceforth
    ``single''):]{Here we take the observation with the highest S/N
    for use as a template and remove it from the observations to be
    timed. This approach has the advantage that all true features of
    the pulse profile are present in the template, but the
    disadvantages that the templates tend to have higher noise levels
    than in alternative approaches and are more susceptible to negative impacts from effects 
    such as scintillation, profile evolution, SWIMS, and man-made extra noise 
    such as mis-calibration or RFI. While this approach is convenient for
    pulsars that have large variations in S/N since the noise present
    in the template for such pulsars may be negligible in comparison
    to the noise present in the typical observation. However, in the case that the dataset is composed of a limited number of observations, it can be disadvantageous to exclude 
    the most valuable (i.e.\ brightest) observation from
    the timing.} 
    
  \item[\textbf{All observations added} (henceforth ``added'') :]{The added 
    template consists of all observations added together, while aligning them using the timing ephemeris 
    of the data. The RFI and misaligned integrations due to, e.g.\ clock offsets 
    are then zapped out. The cleaned archives are then re-weighted with several 
    methods (S/N, S/N$^2$, off-pulse RMS) and the one with the highest resulting S/N is then chosen as the final added 
    template. This
    approach also has the advantage that by definition all true
    profile features are present in the template; and particularly for
    pulsars with stable S/Ns the added template profile would have a
    much higher S/N than the single template. The main disadvantage is
    the risk of self-standarding \citep{hbo05a}, which is due to the fact that noise in 
    observations is correlated with their contributions in the template baseline.
    The impact of this effect is strongly dependent on the number of observations added into the template and on the S/N of the added observations \citep{hbo05a}. Specifically, it has been found to be most relevant for pulsars in which the typical S/N is small (of order $\lesssim 25$) and in cases where less than several hundred observations are added together. The effect is also more problematic for strongly scintillating pulsars since the effective (i.e.\ S/N-weighted) number of added pulses tends to be smaller in this case.}
   
  \item[\textbf{Added with smoothing} (henceforth ``smoothed''):]{Here
    the added profile is taken as a basis and a smoothing filter is
    applied subsequently. Specifically we used
    templates smoothed with the default wavelet-based smoothing
    algorithm of the \textsc{psrchive/psrsmooth} routine \citep{dfg+13}, i.e.\ with an undecimated Daubechies wavelet with
    factor 8. The Daubechies wavelets are a family of orthogonal wavelets
    defining a discrete wavelet transform; the ``undecimate'' version of them is translation invariant; and the factor refers to the number
    of coefficients used \citep{dau92}. Limited testing with variations of the smoothing
    parameters and wavelets used indicated that these generally did
    not result in significantly different results for our
    analysis. The advantage of this approach is that the
    self-standarding issue should be removed as the noise in the
    template is largely suppressed, but this comes at the cost of a
    potential loss or suppression of sharp features in the pulse
    profile. An added complication is that smoothing algorithms can introduce artefacts in the template.  The precise application of the smoothing algorithm could
    in itself be expanded to a far more extensive study, as results
    may also depend on the number of phase bins across a profile
    (i.e.\ the extent to which high-frequency profile features are
    resolved or not), but for the present work we restrict ourselves
    to the most standard application of this method, in order to limit
    the dimensionality of our analysis.}
    
  \item[\textbf{Analytic profile} (henceforth ``analytic''):]{For this
    profile, von Mises functions \citep{js01, ehp01} were fitted to the added
    profile. Given the high S/N of the added profiles, dozens of components 
    were typically required. Components were added and
    fitted interactively to avoid overfitting and to ensure as good a
    template profile as possible. This method has the clear advantage
    that the template profile is truly noise-free, but often not all
    sharp features may be represented in the profile, either because
    they are mis-identified as noise or because the analytic function
    cannot be indefinitely extended without affecting the stability
    and convergence of the least-squares fit. Consequently
    self-standarding is of no concern, but timing precision may be
    compromised.}
\end{description}

An illustration of how the different methods compare is included in
Figure~\ref{fig:ncy_profile} which shows the added templates derived from the NRT data for all three pulsars, as well as the profile differences comparing the added profile to the other three templates. 
As expected, the differences are mostly dominated by white noise, although the added-single difference does show coherent structure, in particular (though not exclusively) in the on-pulse phase range. This is likely caused by calibration issues, scintillation or low-level RFI. 
The added-analytic difference is far better behaved, but here too some non-white signals are visible, demonstrating the limitations of the von-Mises functions in modelling of pulse profile shapes. 
The added-smoothed difference is the closest to pure white noise for all three pulsars, indicating that the wavelet smoothing successfully manages to remove noise and keep structure. 
Nevertheless, even the smoothed template shows a minor deformation near the pulse peak of PSR~J1713+0747. 
To what degree these discrepancies affect the timing of these pulsars, will be assessed in Section~\ref{ssec:templates}.

Note that throughout the presented work the aforementioned templates
are used only in their frequency-averaged form, except for the
analysis focused on the most useful TOA bandwidth, which was based on an added template that was \emph{not} frequency averaged.
A comparative analysis of frequency-resolved analytic templates
\citep[as developed by][]{ldc+14,pdr14} is deferred to a separate
study, given the inherent complexity and multi-dimensional analysis necessary
for such work.

\subsection{TOA determination methods}\label{ssec:CCA}

After creation of a template profile, the TOA of a given observation
can be determined by matching the observation to this
template. Specifically, the observation $\mathcal{O}$ can be
related to the template $\mathcal{T}$ as \citep{tay92}:
\[
\mathcal{O}\left(\phi\right) =
a+b\mathcal{T}\left(\phi+\tau\right)+n\left(\phi\right),
\]
where $\phi$ is the rotational phase, $a$ is an arbitrary offset, $b$
is a scale factor, $\tau$ is the phase offset between observation and
template and $n\left(\phi\right)$ is a noise term. In order to solve
for $\tau$, a variety of possible approaches have been proposed. We
will compare three of the more commonly used methods, described
briefly below.

\begin{description}
  \item[\textbf{Fourier phase gradient (PGS)}:]{The most
    common template-matching method used to date (and the default
    method used in the \textsc{psrchive} software package) is the
    so-called PGS method described in detail by
    \citet{tay92}. Based on the Fourier shift theorem, it matches the
    template to the observation by fitting for a slope in the Fourier
    space. A clear advantage of this approach is that the phase resolution does not impose a fundamental limit on the achievable measurement precision 
    and as such can
    result in significantly more precise measurements than time-domain
    cross-correlation methods \citep[which, according to][are limited
      to a precision about ten times smaller than the data's
      resolution]{tay92}. The main disadvantage of this method is that
    in the low-S/N regime it underestimates the TOA uncertainty since
    the TOA distribution no longer follows a Gaussian
    distribution \citep[see the discussion in][Appendix~B.]{abb+15}}.
  \item[\textbf{Fourier domain with Markov-chain Monte Carlo (FDM)}:]{
    One proposed solution to the underestimation of low-S/N TOA
    uncertainties is to probe the likelihood--phase shift dependence
    with a one-dimensional Markov-chain Monte Carlo, from which the
    TOA variance can be derived. This results in TOA values that are
    identical to those of the PGS method, but TOA uncertainties
    that are more realistic (i.e.\ larger), particularly for low-S/N
    observations.}
 \item[\textbf{Gaussian interpolation shift (GIS)}:]{ This algorithm
  carries out a standard cross-correlation of the template and
  observation in the time domain; and determines the phase offset by
  fitting a Gaussian to the cross-correlation function, whereby the
  centroid of the resulting Gaussian is defined as the TOA; the offset
  required to double the $\chi^2$ of the template-observation
  comparison is defined as the TOA uncertainty \citep{hbo05a}. As
  mentioned above, this time-domain method has limited precision, but
  it was proposed as a more robust TOA determination method in the
  low-S/N regime.
  
  The application of a Gaussian in
  determining the peak position of the cross-correlation function
  should result in timing precision exceeding 10\% of a phase bin,
  although this likely depends on the exact pulse shape.}
\end{description}
\subsection{TOA bandwidth and SWIMS}\label{ssec:TOABW}

The most significant contribution to the TOA uncertainty is referred to as radiometer noise. For a certain telescope, according to the radiometer equation \citep[for details, please refer to Appendix~1 of][]{lk05}, the radiometer noise, $\sigma_{\rm{rm}}$, scales with bandwidth $B$, integration time $T$ as,
\begin{equation}
\label{eq:RM}
\sigma_{\rm{rm}} \propto \frac{1}{\sqrt{TB}}.
\end{equation}

In general, the sensitivity of pulsar timing would increase with the increase of observational bandwidth. However, as the TOA bandwidth increases, other noise sources become increasingly significant and the traditional approach of deriving TOAs from fully frequency-averaged observations can become increasingly complex, particularly if scintillation and profile evolution are significant.

At the other extreme, TOAs determined using native frequency resolution (referred to hereafter as ``fully frequency-resolved TOAs'') significantly decreases the S/N, possibly leading to 
the low-S/N regime where some template-matching methods don't
provide reliable uncertainties anymore; in addition, TOAs at native frequency resolution can substantially increase the data volume and hence the computational cost of timing analyses, particularly for wideband systems. Consequently, there must be an most useful TOA bandwidth which maximises the TOA precision by limiting the deleterious effects of low S/N in fine frequency channels and those due to scintillation and profile evolution in frequency-averaged TOAs. 

Here we have investigated such an most useful TOA bandwidth for a
set of pulsar and backend combinations, by quantifying the achievable
timing precision as a function of the TOA bandwidth. To do so, we
combined the best template-matching method with a
frequency-resolved version of the optimal template profile, and then 
carried out the timing of each pulsar and backend combination of
our test dataset for a range of possible TOA bandwidths. This 
allowed us to investigate the improvement in the goodness of fit (of the timing model to the TOAs) via the reduced
chi-squared of the linearised least-squares fit and the RMS of the timing
residuals, as well as the number of TOAs that remain after removing TOAs
corresponding to total intensity profiles with S/N less than 8, following
\citet{abb+15}.

\begin{figure}
 \includegraphics[width=\columnwidth]{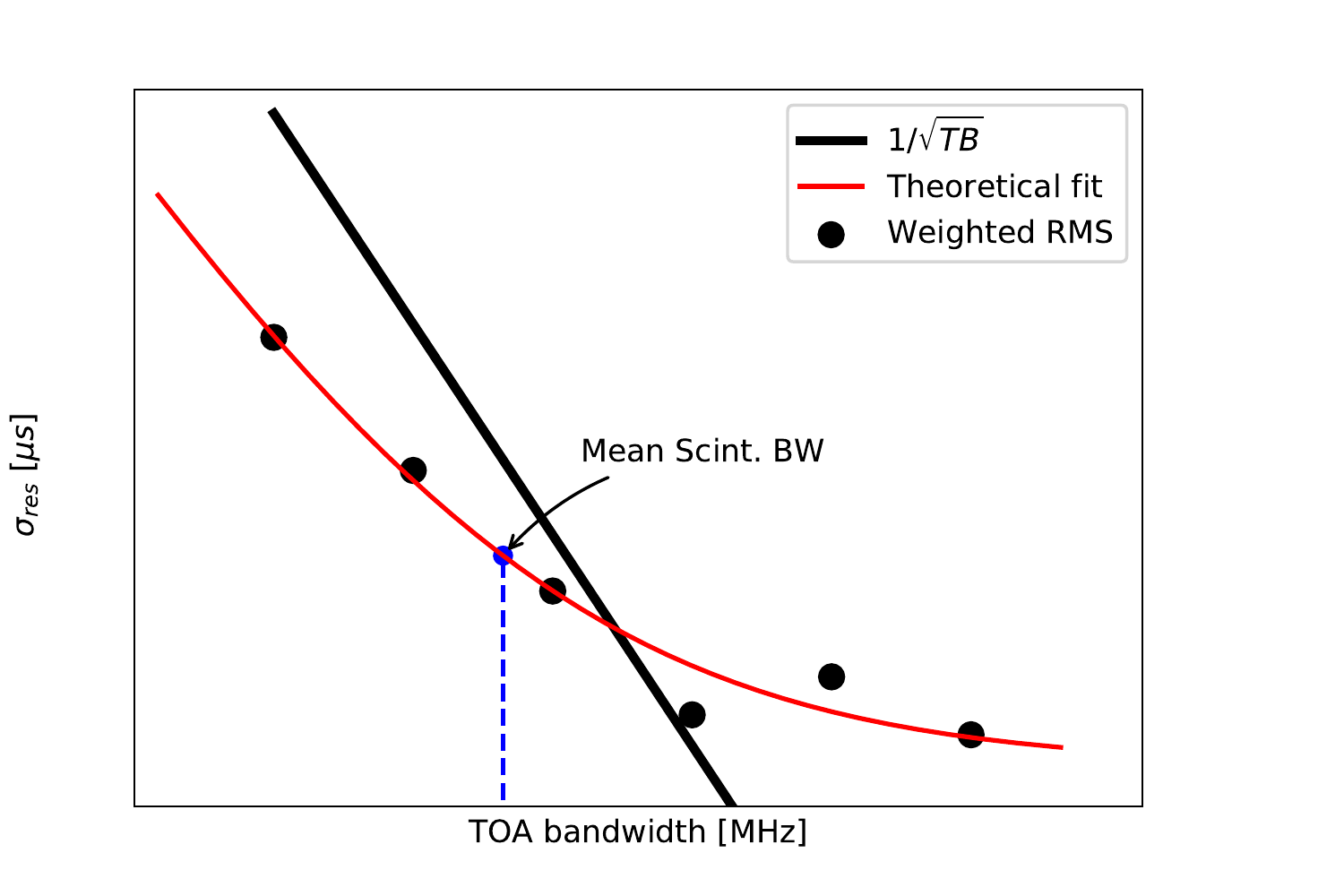}
 \caption{A sketch of the dependence of $\sigma_{\text{res}}$ on TOA bandwidth, for an idealised dataset. The black dots represent RMS
   values, the blue dashed line indicates the median scintillation
   bandwidth and the red line is a fit of
   Equation~\ref{eq:WN}. The black line is for the case of pure radiometer noise, i.e.\ with no jitter or other corrupting noise sources ($\sigma_{\rm NF} = 0$) and for
   standard scaling following the radiometer equation ($\alpha = -1$)}
 \label{fig:J1713_toabw}
\end{figure}

Radiometer noise is proportional to the square root of the TOA bandwidth
and jitter noise or SWIMS \citep{ovh+11,cd85, lkl+12} behaves as a constant noise term, added in
quadrature to the radiometer noise. The combined effect of profile evolution and scintillation strongly depends on the specific pulsar properties. Theoretically this effect should get stronger with larger TOA bandwidths, but given the relatively narrow bandwidth of our data, it is likely to behave somewhat differently. In addition, a number of systematic effects \citep[e.g.\ calibration errors or instrumental artefacts; see][for a full review]{vs18} are not expected to strongly depend on TOA bandwidth, although this would again be highly variable and dependent on the particular situation. In analysing the timing residual RMS as a function of TOA bandwidth, we consequently do not expect the straightforward scaling of the radiometer noise (Eq.~\ref{eq:RM}), but expect a pulsar-dependent slope and likely some form of RMS saturation at the widest bandwidths -- we will henceforth refer to the level of this RMS saturation as the ``system limited noise floo'' or SLNF. Consequently, we will quantify the residual RMS as a function of TOA bandwidth as follows: 
\begin{equation}\label{eq:WN}
\sigma_{\text{res}} = \sqrt{\frac{CB^{\alpha}}{T}+\sigma_{\text{NF}}^2},
\end{equation}
where $\sigma_{\text{res}}$ is the residual RMS of the post-fit
residuals, $C$ is a constant,
$T$ is the integration time of the TOAs, $B$ is the bandwidth of the TOAs, $\alpha$ is a scaling index which would be $-1$ in the case of pure radiometer noise 
and $\sigma_{\text{NF}}$ is 
the SLNF, i.e.\ the part of the RMS that does not scale with TOA bandwidth. This last term is mostly expected to quantify the impact of SWIMS\footnote{For the relatively limited fractional bandwidths
considered in this work, the impact of SWIMS cannot be expected to depend on TOA bandwidth 
\citep{sbh+99}.} but also comprises other effects as described above.  An example of this dependence is shown in Figure~\ref{fig:J1713_toabw}.

While the above considerations prevent a meaningful connection to the
physical phenomenon of SWIMS, our analysis does allow a
phenomenological determination of which TOA bandwidth optimally
balances the various effects that impact the final timing RMS. 

\section{Simulations} \label{sec:simulation}
Since the analysis on real data is complex and multi-faceted, we first carry out some relatively straightforward simulations to inform us on a few key points of the subsequent analysis. Specifically, we use simulations to study how the TOA uncertainties from the various TOA determination methods scale with pulse S/N; and to check to what degree these results depend on the pulse shape. Furthermore, since these tests depend on the S/N, which is being used to place a cut-off to remove unreliable TOAs \citep[in accordance with the advice of][]{abb+15}, we also evaluate how well the standard S/N algorithm of the \textsc{psrchive} package can reproduce the simulated S/N. 

\begin{figure*}

\begin{minipage}[h]{1\linewidth}
\centering
\subfloat[Simple Gaussian profile.]{\includegraphics[width=.3\linewidth]{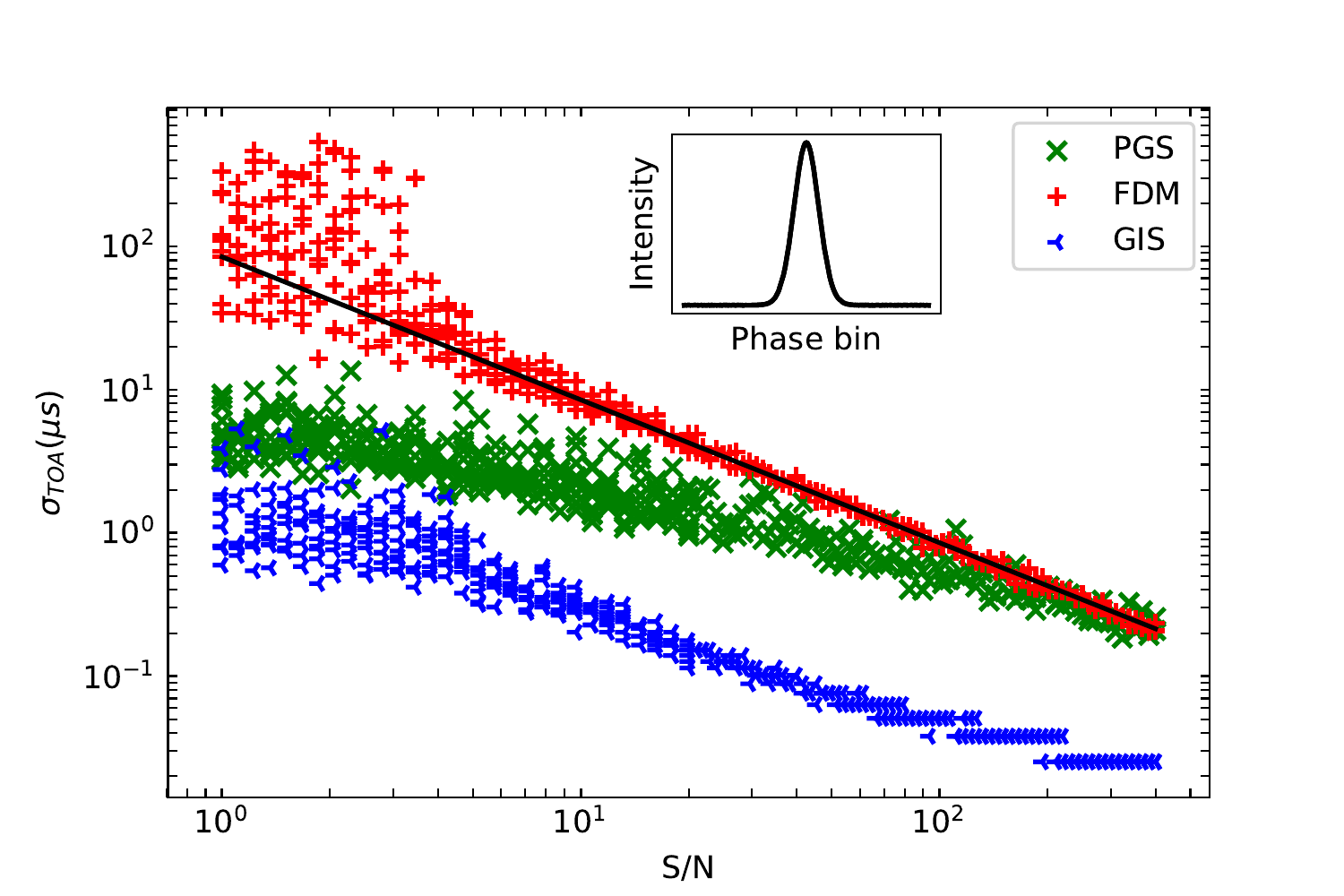} \label{fig:simulation01}}
\quad
\subfloat[Gaussian profile with a narrow notch.]{\includegraphics[width=.3\linewidth]{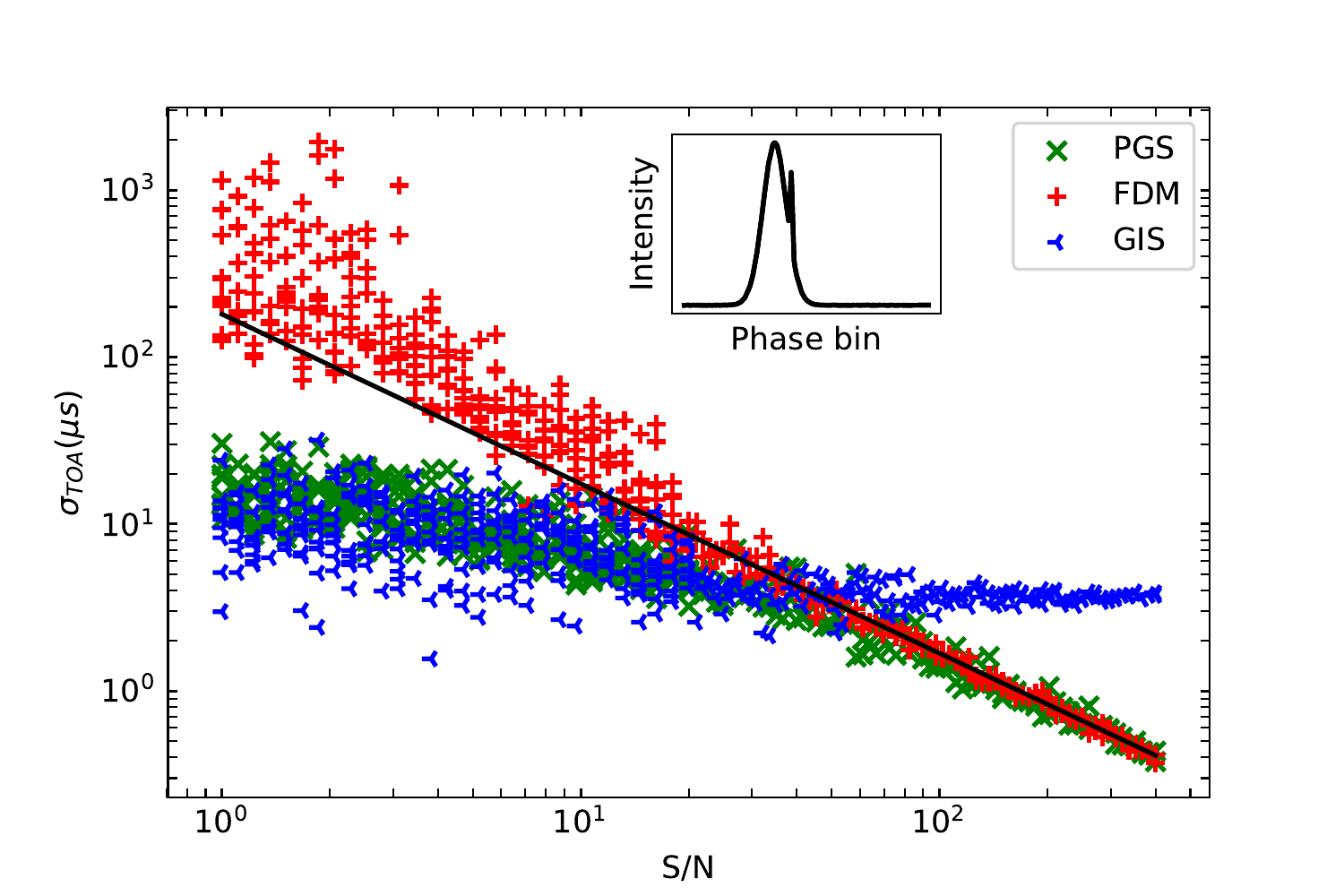}\label{fig:simulation02}}
\quad
\end{minipage}

\begin{minipage}[h]{1\linewidth}
\subfloat[PSR~J0218+4232.]{\includegraphics[width=.3\linewidth]{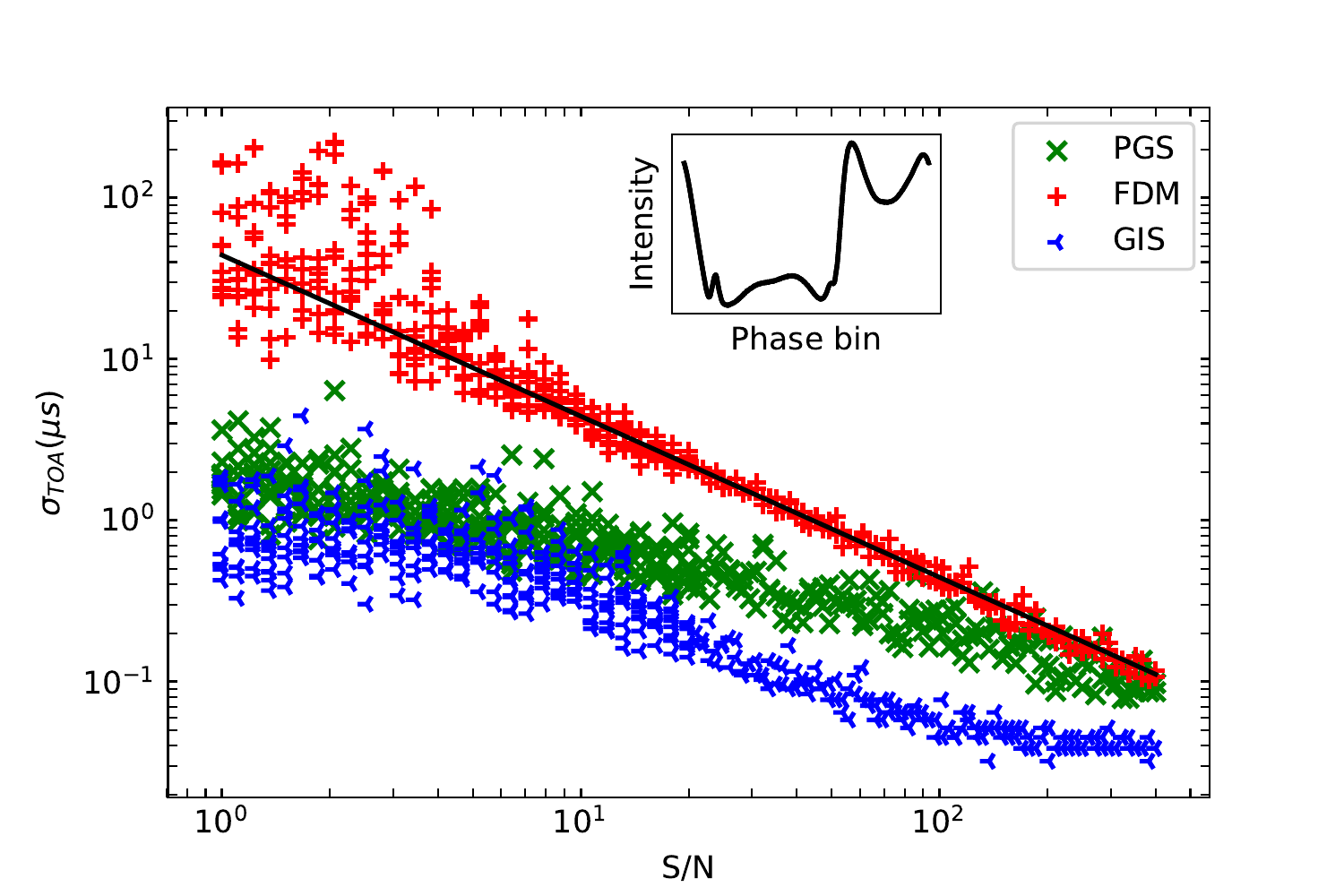}\label{fig:simulation0218}}
\quad
\subfloat[PSR~J1713+0747.]{\includegraphics[width=.3\linewidth]{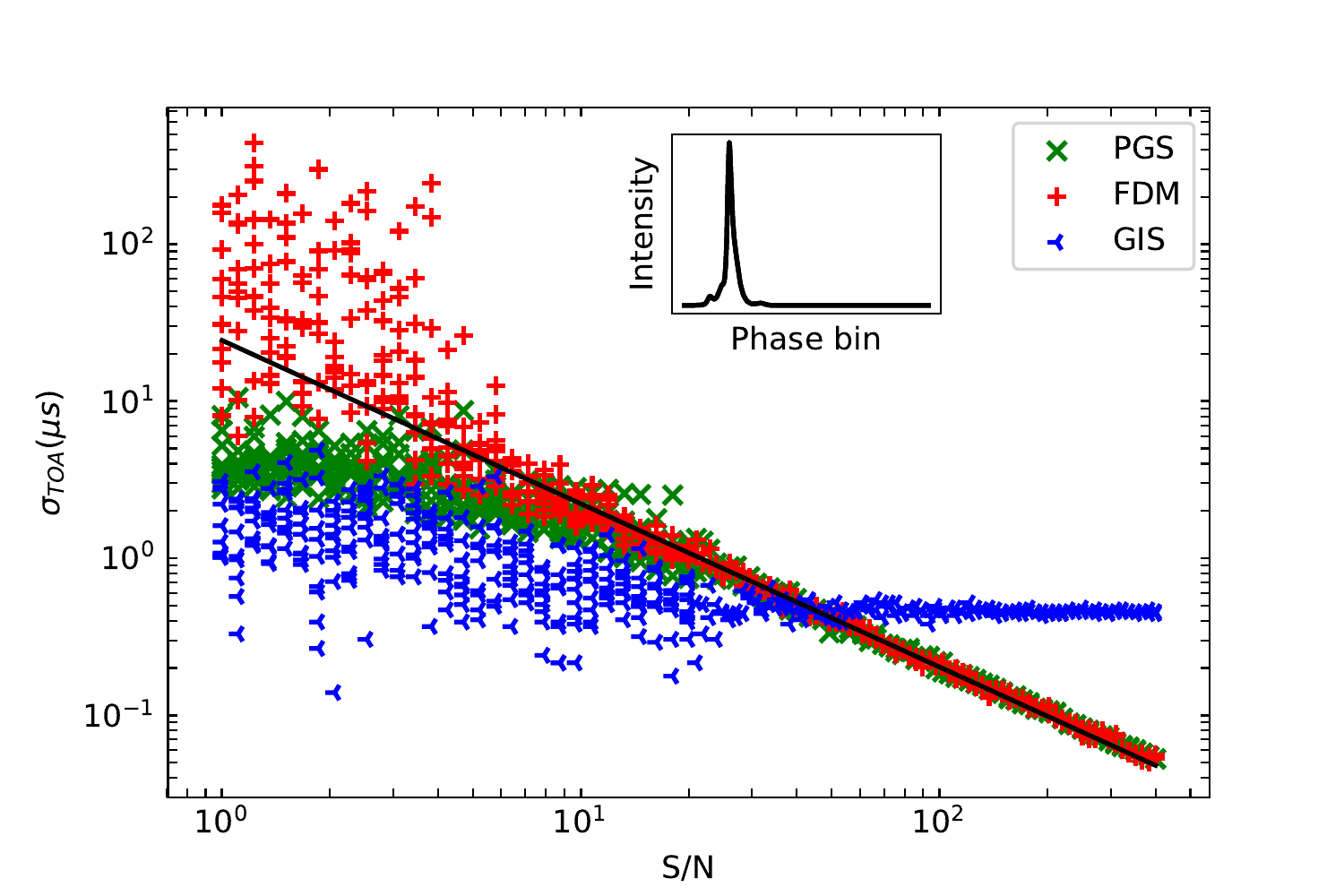}\label{fig:simulation1713}}
\quad
\subfloat[PSR~J2145$-$0750.]{\includegraphics[width=.3\linewidth]{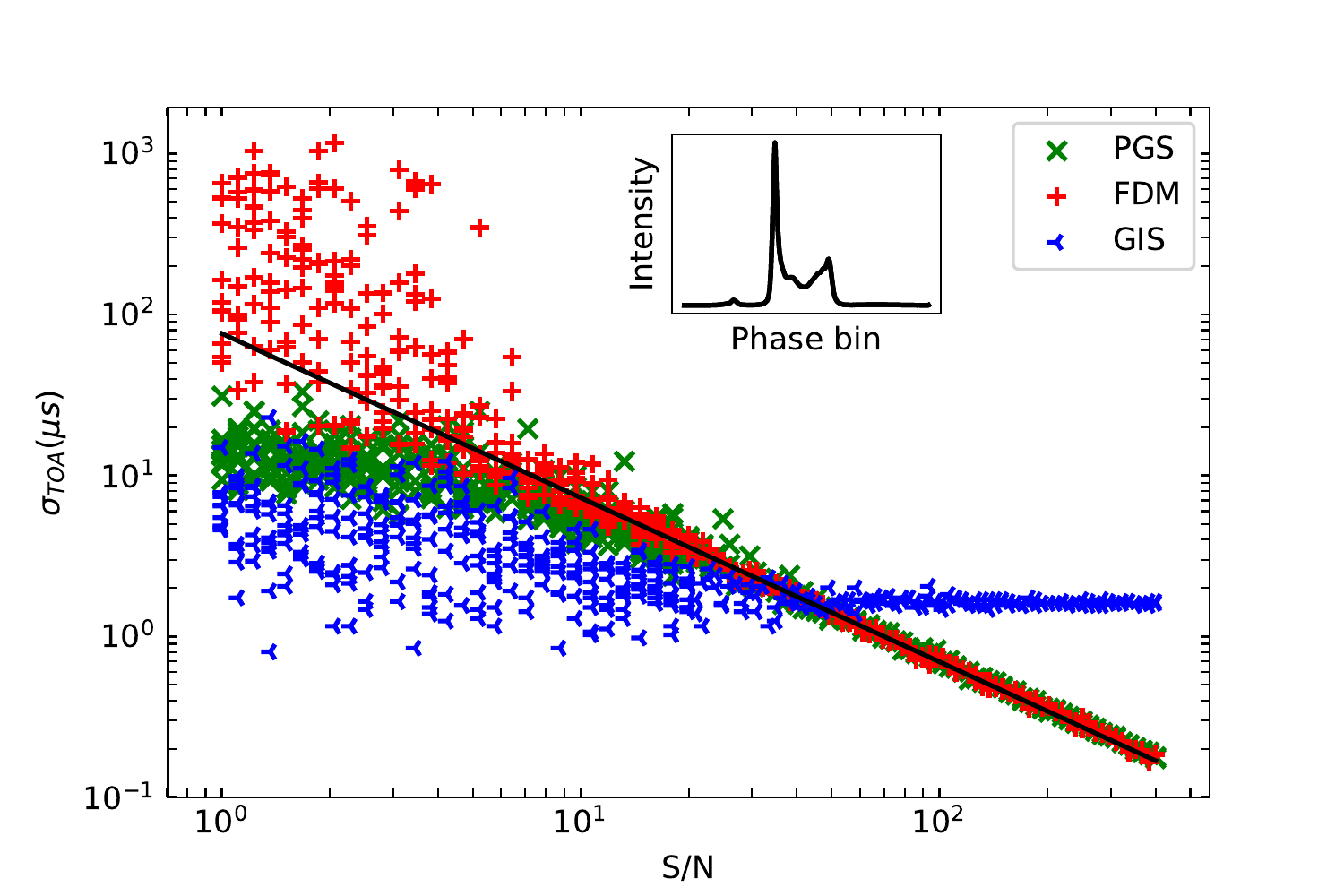}\label{fig:simulation2145}}
\end{minipage}

\caption{TOA uncertainty obtained from various noise-added profiles as a function of the simulated S/N. The top-left panel shows the results for a simple Gaussian profile, the top-right panel shows results for a Gaussian profile with a notch, and the bottom panels display the results for the analytic templates of PSRs~J0218+4232, J1713+0747, and J2145$-$0750, respectively, as derived from NCY data. CCAs are shown with different colours and markers, as indicated in the legend. The black solid line in each panel is a fit to high-S/N (> 10) FDM data and is extended to low-S/N data. At each S/N plotted, ten simulations were run, although these are typically indistinguishable at high S/N. The inset within each panel shows the profile.}
\label{fig:simulation}
\end{figure*}

These simulations were written in \textsc{python} using the \textsc{numpy} package and \textsc{psrchive} through its python interface. Specifically,  noise-free templates were constructed with the \textsc{paas} routine of \textsc{psrchive}, after which white noise with varying intensity was added to those profiles. Here we tested five profile shapes as shown in the inset plots of Figure~\ref{fig:simulation}: a simple Gaussian profile; a Gaussian profile with a sharp notch on its trailing edge; and finally the analytic profiles derived for each of our three test pulsars, based on the NCY data described earlier.

Ideally, according to the radiometer equation, the TOA uncertainty should follow an inverse relationship with S/N, in the form of: $\sigma \propto \frac{1}{S/N}$. From Figure~\ref{fig:simulation}, we can see that PGS and FDM are consistent with each other in most cases and follow this scaling in the high-S/N regime, although a minor deviation in the slope of the PGS curve can be identified for PSR~J0218+4232. However, if the S/N is low, i.e.\ less than $\sim$10, the difference between FDM and PGS becomes more explicit, although the exact threshold S/N strongly depends on the pulse shape: for the complex profile shapes of PSRs~J1713+0747 and J2145$-$0750 the curves coincide down to lower S/Ns; for the simple Gaussian or Gaussian-with-notch profiles the discrepancies are clearer even at S/N=10. These results clearly show that PGS typically underestimates the TOA uncertainties in the low-to-medium S/N regime, although the degree of underestimation strongly depends on the profile shape. In contrast, FDM displays a tendency to \emph{over}estimate the TOA uncertainties in the same S/N regime, primarily for pulse profiles with sharp features. 

It is furthermore evident that GIS fails to determine the uncertainty correctly in both the low-S/N and high-S/N regimes, except for a very narrow region. Similar to the results from \citet{hbo05a}, our simulations show that GIS works better at the low-S/N regime; and the range of good performance is wider for pulsars with a large duty cycle. Nevertheless, our simulations suggest the uncertainties are consistently underestimated, even at low S/Ns. This latter difference may be partly related to the S/N of the template used, which we made unrealistically large to more clearly demonstrate the trends independently of template; although clearly the complexity of the pulse profile shape also has a significant impact, as already anticipated by \citet{hbo05a}. For profiles with narrow features, our simulations for GIS show that the TOA uncertainty flattens off for S/Ns above a few tens, indicating significant overestimation of uncertainties for high-S/N observations.

\begin{figure}
    \centering
    \includegraphics[width=\columnwidth]{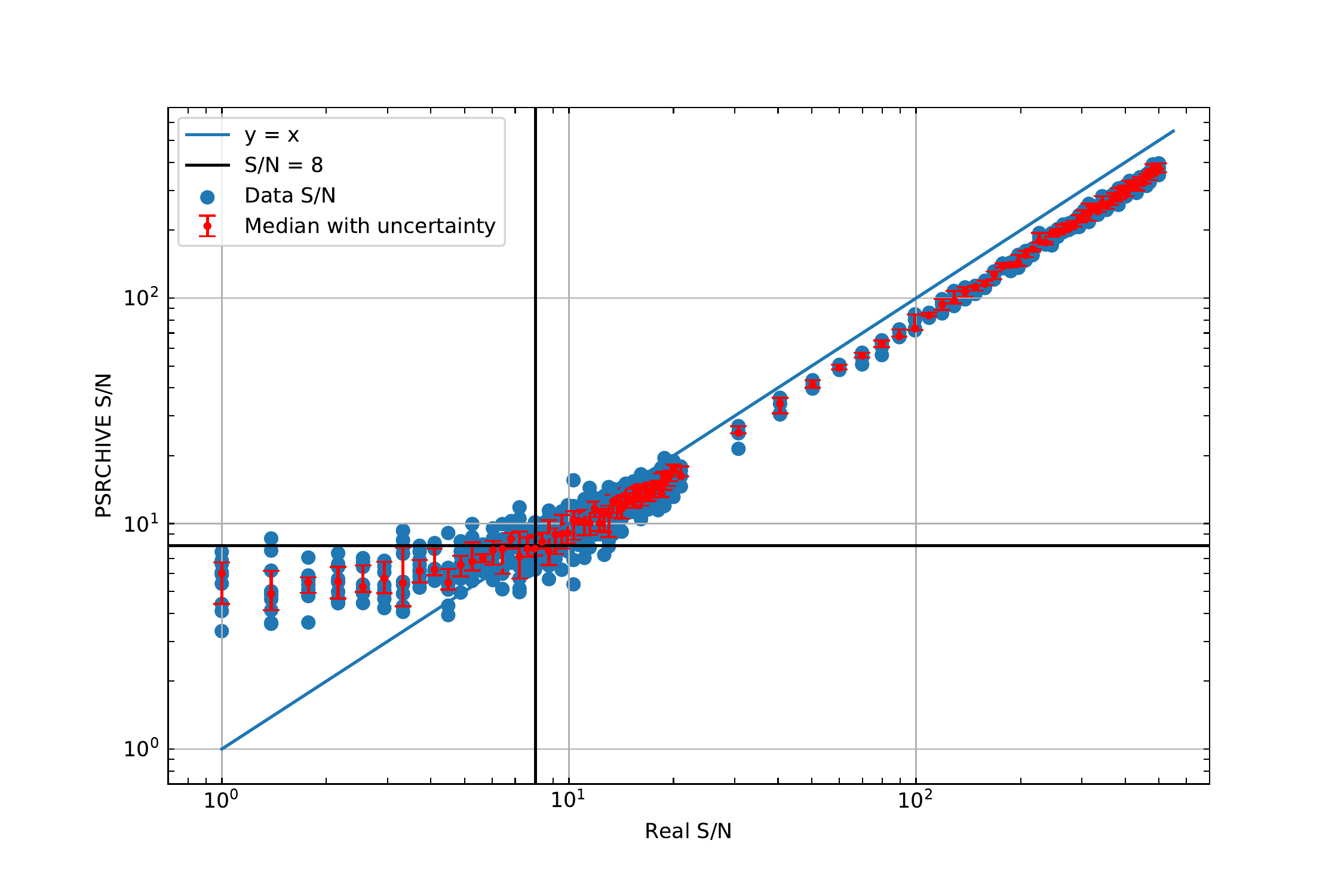}
    \caption{S/N determined with the standard implementation in \textsc{psrchive} as a function of simulated S/N. The blue points are the S/N of the individual simulations, while the red points indicate the median value of the simulations, the red error bars indicate the 25th and 75th percentiles at each real S/N. For this simulation, the analytic template based on NRT data of PSR~J2145$-$0750 was used. Due to differences in the definition of the S/N, some inconsistencies between x and y values are expected, although the scaling should be linear, as is the case in the high-S/N regime. In the low S/N regime (S/N$<10$), the S/Ns returned by \textsc{psrchive} are typically overestimated and have a significant random component. (The thick black lines indicate S/Ns of 8, which is used as a S/N cut-off value in this work, following \citet{abb+15}.)}
    \label{fig:SNevol}
\end{figure}

Figure~\ref{fig:SNevol} shows the results of our consistency check on the S/N values returned by \textsc{psrchive}. Since the exact S/N depends on the precise definition used as well as on the pulse shape, no exact equivalence can be expected between the simulations and the measurements, but a linear relationship should exist. Our results show that this linear relationship does indeed describe the majority of the simulated S/N range, but that this flattens off below a S/N of $\sim$10, where the S/N returned by \textsc{psrchive} is mostly unrelated to the actual S/N simulated. While the details of this figure -- and particularly the scaling at high-S/N -- differ between the various pulse profiles we tested, the flattening off below a S/N of ten is persistent. 

A consequence of this test lies in the interpretation of Figure~\ref{fig:simulation}: while for some pulsars the TOA uncertainties from FDM and PGS appear reliable down to a S/N of approximately five, in practice it is not possible to reliably measure S/Ns below ten. Consequently, in order to reliably identify TOAs with reliable error bars, a cut-off at a S/N of about ten may be required in any case, which could cause any difference between PGS and FDM to become irrelevant for a wide range of pulse shapes. (In contrast, Figure~\ref{fig:simulation0218} indicates that for pulse shapes with particularly large duty cycles differences may persist even above a S/N of ten.)

\section{Results and discussion}\label{sec:results} 

To evaluate the timing precision as a function of the choice of
template and CCA described in Sections \ref{ssec:std} and \ref{ssec:CCA}, 
and to study the most useful TOA bandwidth as described in Section~\ref{ssec:TOABW}, we
analysed the TOAs with the timing models presented by \citet{dcl+16} and \citet{ccg+21}. For PSR~J1713+0747 we adopted the entire timing model of \citet{ccg+21}, including DM and red noise models. For PSRs~J0218+4232 and J2145$-$0750 we used the timing models from \citet{dcl+16}, but without inclusion of the red noise or DM models. In the timing models of \citet{dcl+16} the planetary ephemerides were updated to DE438 and the reference clock was updated to TT(BIPM2019).

\subsection{Template and CCA}\label{ssec:templates}

For each combination of the templates and CCAs described in Section
\ref{sec:proc}, we performed a \textsc{tempo2} \citep{ehm06, hem06} analysis,
fitting only for the pulsar spin and its first derivative. For each fit, the
reduced $\chi^{2}$ and the residual RMS value were recorded, along with the resulting timing residuals.

\begin{figure*}[htbp]
\centering
\includegraphics[width=0.9\textwidth]{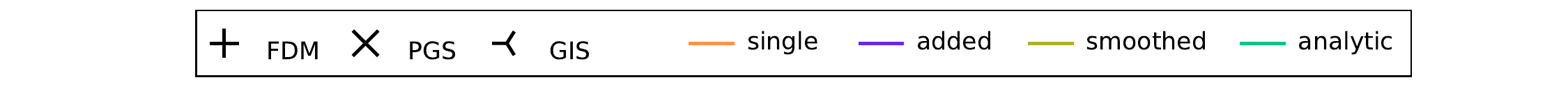}
 \includegraphics[width=0.9\textwidth]{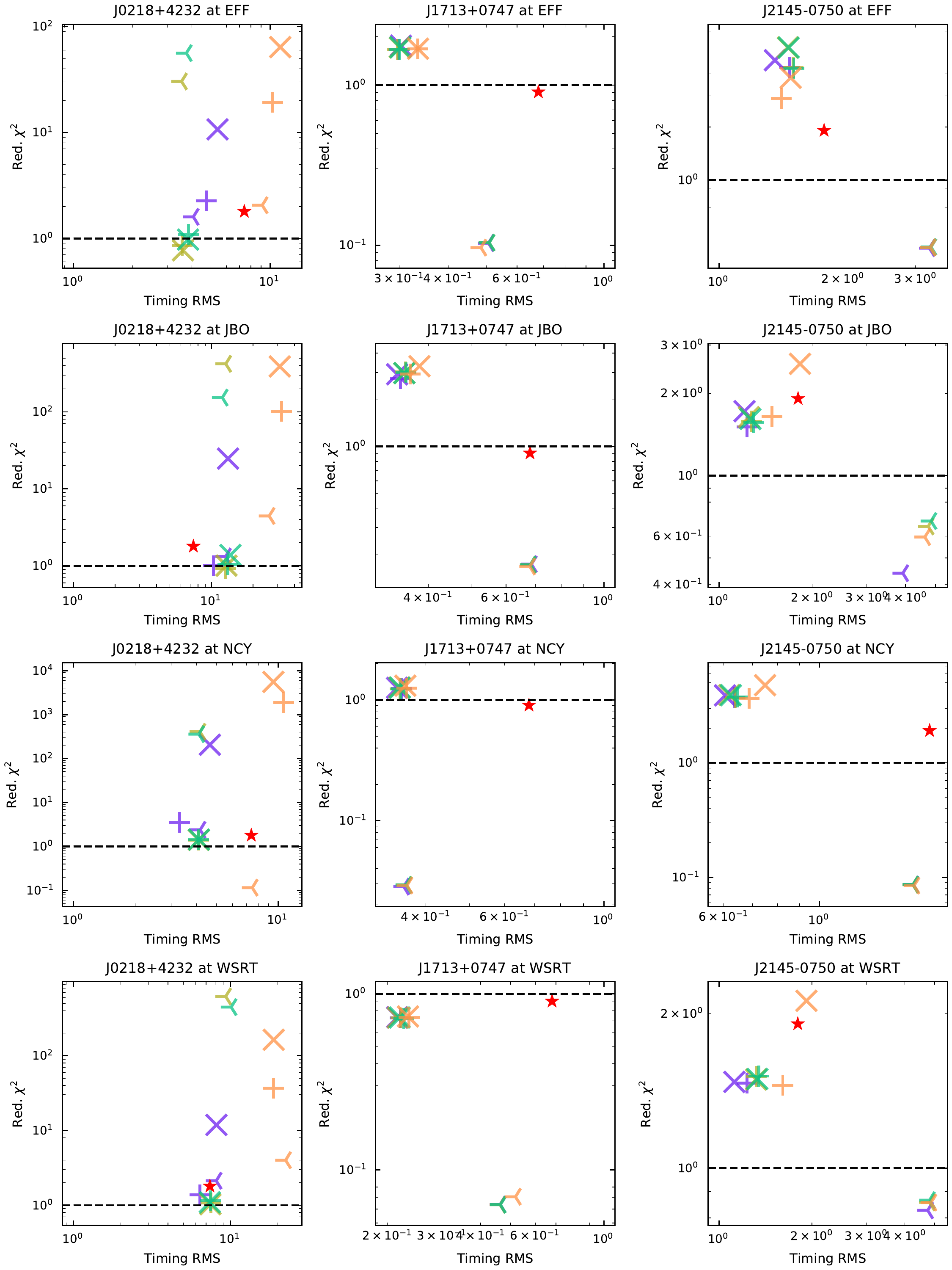}
\caption{Reduced $\chi^2$ and residual RMS for the three pulsars as a function of the
  selected CCA and template and the different telescopes. The dashed line indicates a reduced $\chi^2$ of 1, and the red stars represent results from \citet{dcl+16}}\label{fig:chiVrms}
\end{figure*}

Figure~\ref{fig:chiVrms} presents a pulsar and telescope-wise analysis of the timing performance. Here the poor performance of the GIS TOAs is clearly visible, across all systems. Further, it is evident that FDM and PGS results often cluster for the two stronger sources, with generally almost identical reduced $\chi^2$ and RMS values, independent of the template creation method. This shows that the effect of template choice can be somewhat mitigated for such sources when these CCAs are used. Finally, for those cases where FDM and PGS do differ, FDM tends to outperform PGS, particularly in terms of the reduced $\chi^2$ values obtained.

In Figure~\ref{fig:chiVrms}, it is also noticeable that for faint pulsars like PSR~J0218+4232, the high noise levels in the single template severely compromise the results. In this case, even the added template shows poor results when combined with the PGS method, indicating that some level of self-standarding \citep[as described by][]{hbo05a} may be at work. However, the FDM method appears far more robust to this phenomenon.

The timing residuals for each of the TOA sets were further assessed to check for the presence of unmodelled signals or persistent outlier TOAs. Using a 2-tailed Kolmogorov-Smirnov test we compared the power spectrum of these residuals against a power spectrum consisting of purely Gaussian or power-law signals or a combination of both. We find that the power spectra of the timing residuals for all three pulsars are Gaussian or a Gaussian and power-law mixed distribution, suggesting that the updated timing models are sufficient given our data (i.e.\ the timing-model parameters that were not refitted model the TOAs sufficiently well that no clear timing signatures were present in the data). Visual inspection of the timing residuals confirms that the timing residuals of PSRs~J0218+4232 and J2145$-$0750 are essentially white; the timing residuals of PSR~J1713+0747 still show some timing-noise signal that is not sufficiently modelled, this feature is mostly a consequence of a so-called "event" which is known to be hard to model with standard red-noise models \citep{leg+18, abb+20}. 

The remaining structure in the TOAs of PSR~J1713+0747 results in some non-unity reduced-$\chi^2$ values, although the temporal nature of this feature causes different telescope sub-sets to be affected to different degrees. The significantly elevated reduced-$\chi^2$ values in the PSR~J2145$-$0750 Effelsberg and Nan\c cay data are likely due to a few highly precise TOAs that are affected by the Solar Wind \citep{tsb+21}.

As a reference point for comparison, the reduced-$\chi^2$ and RMS values obtained by \citet{dcl+16} are shown by a red star symbol. Even though their analysis was based on a combined analysis of all four telescopes, spanned a different date range, used data from older observing systems and fitted for all timing-model parameters, our results largely agree with this, although our timing precision tends to be slightly better, as expected.

\subsection{TOA bandwidth}
For each pulsar, the timing performance as a function of the TOA bandwidth was assessed using TOAs created with the `added' template with the FDM CCA, by splitting the archives into multiple channels. (In doing so, the template was split into the same number of channels, implying that each channel was timed against a template at its own frequency.) The exact choice of the number of channels is specific to the telescope to accommodate for the different bandwidths. At EFF, the NRT and WSRT the fully frequency-resolved archives were 
first averaged down to 32 channels and subsequently averaged down by factors of two.  
Due to
the different native resolution at Jodrell Bank, the JBO data were first averaged down to 40 channels (or less) and subsequently averaged down by factors of either 2.5 or two. 

\begin{figure}[htbp]
\begin{subfigure}[b]{\linewidth}
    \centering
    \includegraphics[width=\hsize]{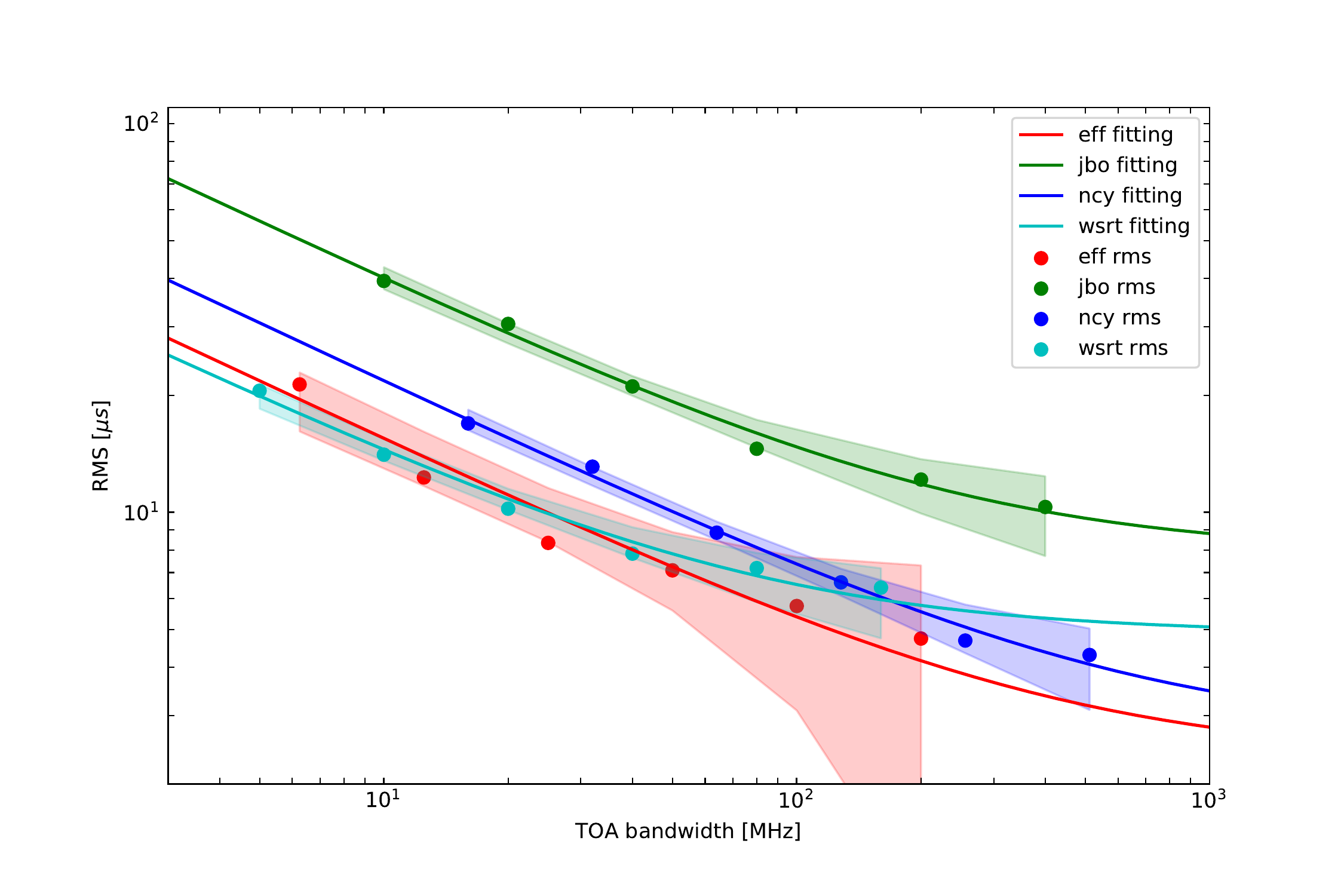}
    \vspace*{-2\baselineskip}
    \caption{`System limited' noise floor estimates for PSR~J0218+4232.}
    \label{fig:j0u_toabwa}
    \vspace*{-1\baselineskip}
\end{subfigure}
\begin{subfigure}[b]{\linewidth}
    \centering
    \includegraphics[width=\hsize]{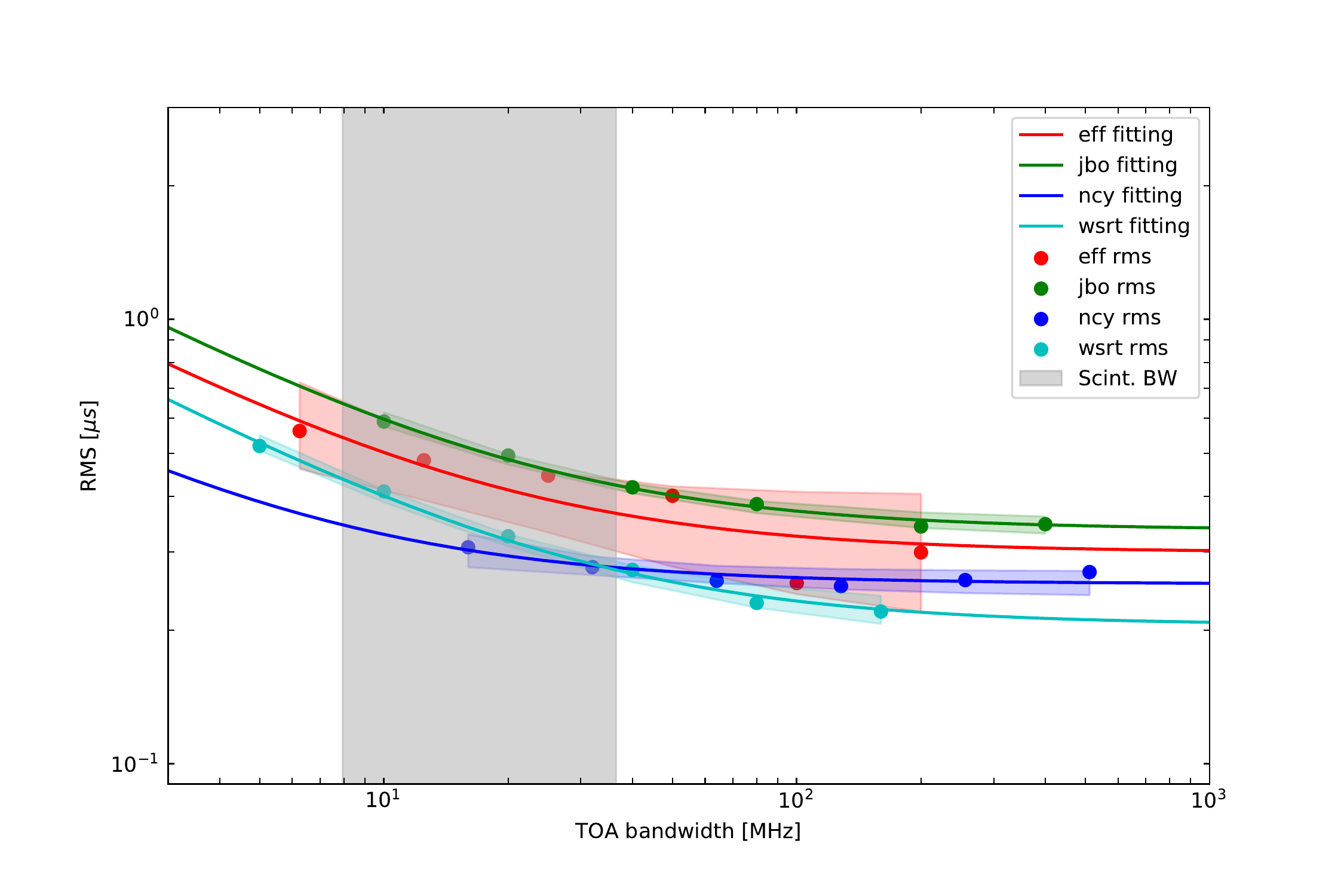}
    \vspace*{-2\baselineskip}
    \caption{`System limited' noise floor estimates for PSR~J1713+0747.}
    \label{fig:j1u_toabwb}
    \vspace*{-1\baselineskip}
\end{subfigure}
\begin{subfigure}[b]{\linewidth}
    \centering
    \includegraphics[width=\hsize]{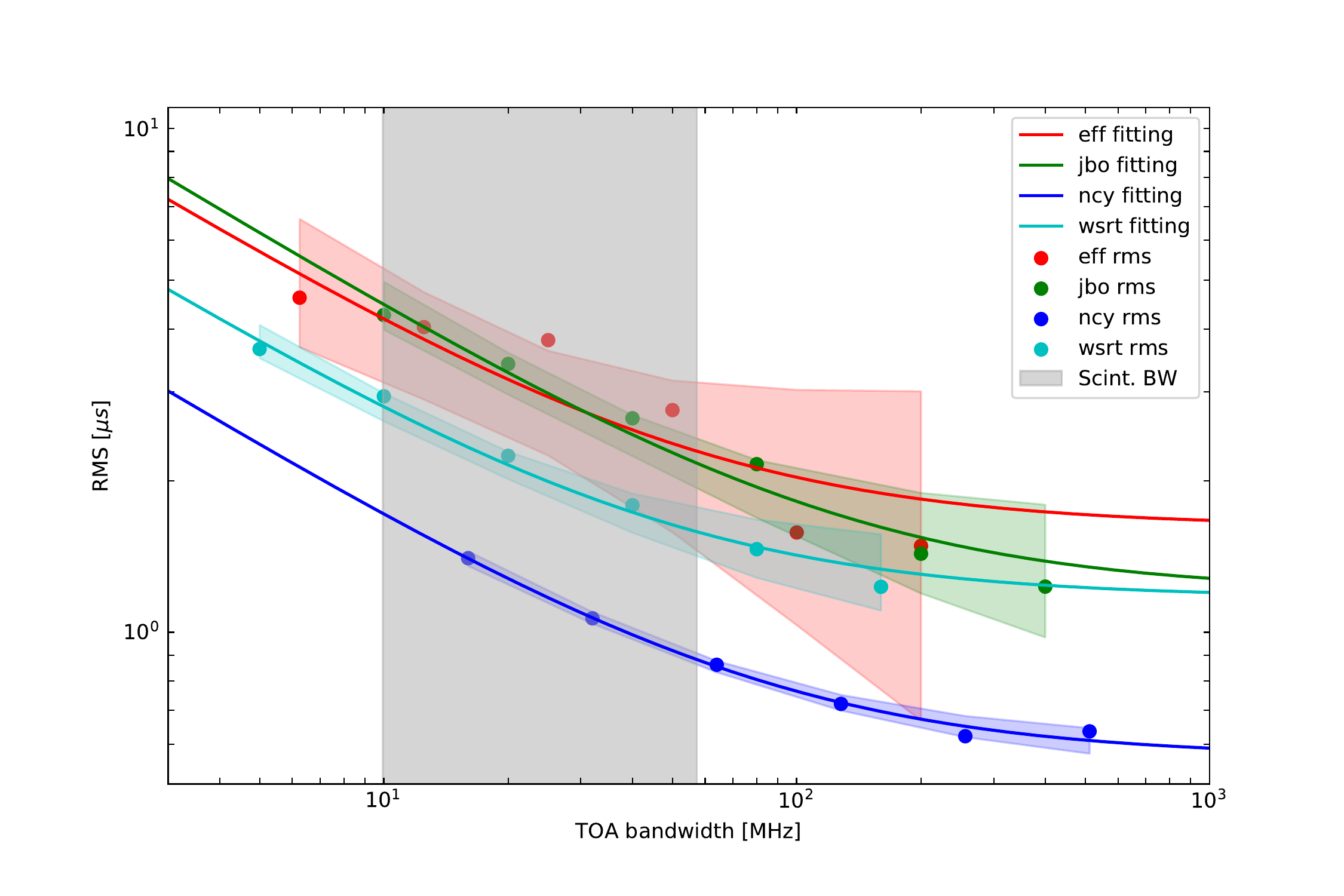}
    \vspace*{-2\baselineskip}
    \caption{`System limited' noise floor estimates for PSR~J2145$-$0750.}
    \label{fig:j2u_toabwc}
\end{subfigure}
\caption{System limited noise floor (SLNF) estimates for the three pulsars, when using
the added template with the FDM CCA to generate TOA with the available bandwidth
at each telescope divided into the respective number of channels. The shaded
regions along the fitted curves are the interquartile ranges for the fit,
representing the error bounds on the estimated SLNF. The grey vertical bands
show the expected scintillation bandwidth, estimated from the same dataset as we present here \citep{lvm+21}.} 
\label{fig:toabw} 
\end{figure}

Following the method described in Section~\ref{ssec:TOABW}, for each choice of number of channels, a standard \textsc{tempo2} timing analysis was carried out, fitting for the same timing parameters as in Section~\ref{sec:proc} (i.e.\ pulse period and period derivative), along with DM and DM derivatives for frequency-resolved data. The resultant RMS for each pulsar and telescope is plotted in Fig.~\ref{fig:toabw}.

From these plots, it is evident that the telescope dependent, achievable RMS decreases asymptotically as a function of the TOA bandwidth, down to the SLNF, as shown by the fit lines in Fig.~\ref{fig:toabw}. A combination of factors influence this limit on the minimum RMS.

For PSR~J0218+4232, where the overall S/N of the profile continuously increases with greater channel bandwidth, the overall timing RMS also continuously decreases. There does appear to be some sign of flattening, so beyond $\sim500$ MHz, the RMS may start to approach the SLNF.

Although PSR~J0218+4232 is known to exhibit ``Large Amplitude Pulses'' \citep{jkl+04}, we do not find a significant influence of those pulses in our data because of the extended integration lengths which imply averaging over several million pulses for each TOA.

For PSRs\ J1713+0747 and J2145$-$0750, we find clearer signs of flattening in the RMS curves, indicating that fully averaging the TOAs across bandwidth may be unwise in these cases. Specifically very little gain is made when averaging over more than $\sim$ 100 MHz TOA bandwidth. 

As described in Section \ref{ssec:TOABW}, these SLNFs include a variety of effects, including system and band noise
as defined in e.g.\ \citet{lsc+16} and effects due to scintillation. Consequently it could be considered that the most useful TOA bandwidth may depend on the scintillation bandwidth of the pulsar in question. However, this does not appear to be the case for our sample -- PSR~J0218+4232 has a scintillation bandwidth well below our finest resolution; while PSRs~J1713+0747 and J2145$-$0750 have bandwidths that are probed by our analysis, but the behaviour of the curves does not significantly differ near the bandwidth range of the scintillation. A full investigation of the SLNF is beyond the scope of this paper, but would be in line with earlier analyses like those by \citet{lcc+16} and \citet{dnc+18}.

\section{Conclusions}\label{sec:conc}

In this paper we investigated the relative merits of three common TOA determination methods and four ways to generate timing templates, by means of comparing their impact on the timing residuals and reduced $\chi^2$ value for data sets from four different telescopes on three different pulsars. In addition, we checked how the timing RMS reaches a plateau as TOA bandwidth is increased. This plateau or system-limited noise floor (SLNF) consists of a variety of factors, including SWIMS, instrumental noise, timing noise etc. and we have shown that it depends strongly on the telescope and pulsar in question.

With regards to the choice of the template generation scheme, we find that the
single brightest observation leads to the worst timing performance. For the
brighter pulsars the added, smoothed and analytic templates lead to
comparable, minimum timing RMS and red. $\chi^2$, but for the lower S/N
PSR~J0218+4232 added templates can perform worse than the smoothed and
analytical templates.

The analysis presented above also upholds the recommendation of \citet{vlh+16} that
the CCA of choice for high-precision PTA work is the FDM method, although the PGS method turned out to be similarly reliable. In
all pulsar, telescope and backend combinations tested above, we find that FDM and PGS 
TOAs lead to the most reliable TOAs and TOA uncertainties, although in the low-S/N regime both methods suffer systematic issues. 
Given our finding that PGS TOAs may be as reliable as FDM-based ones, particularly for bright pulsars, a re-analysis of archival data for which PGS TOAs may already be available, may well be unwarranted. In some cases,
particularly when noisy templates are used, FDM does clearly outperform PGS in
terms of the reliability of TOA uncertainties.

In addition, we find that GIS-derived TOAs are not suitable for high-precision
timing, leading to conservative timing models and a loss of model parameter sensitivity.

Using the FDM CCA with added templates, we find that the most useful TOA bandwidth
for minimising the achievable timing residual RMS is mostly dependent on the pulsar brightness and instrumental sensitivity. For the sample studied here, fully frequency-averaged TOAs seem advantageous for the relatively faint PSR~J0218+4232, whereas significantly narrower bandwidths seem optimal for the brighter two pulsars in our sample.

\begin{acknowledgements}

  The EPTA is a multinational European collaboration, which consists
  ASTRON (NL), INAF/Osservatorio Astronomico di Cagliari (IT),
  Max-Planck-Institut f\" ur Radioastronomie (GER), Observatoire de
  Paris/Nan{\c c}ay(FRA), the University of Manchester (UK), the
  University of Birmingham(UK), the University of Cambridge (UK) and the University of Bielefeld (GER). 

 JW acknowledges support from the China scholarship council. J.P.W.V.\ acknowledges support by the Deutsche Forschungsgemeinschaft(DFG) through the  Heisenberg programme (Project  No.\ 433075039).
 Part of this work is based on observations obtained the 100-m telescope of the MPIfR (Max-Planck-Institut f\"ur Radioastronomie) at Effelsberg. Pulsar research at the Jodrell Bank Centre for Astrophysics and the observations using the Lovell Telescope is supported by a consolidated grant from the STFC in the UK. The Nan\c{c}ay Radio Observatory is operated by the Paris Observatory, associated with the French Centre National de la Recherche Scientifique (CNRS). We acknowledge financial support from the Action F\'ed\'eratrice PhyFOG funded by Paris Observatory and from the ``Programme National Gravitation, R\'ef\'erences, Astronomie, M\'etrologie'' (PNGRAM) funded by CNRS/INSU and CNES, France. The Westerbork Synthesis Radio Telescope is operated by the Netherlands Institute for Radio Astronomy (ASTRON) with support from The Netherlands Foundation for Scientific Research NWO.

\end{acknowledgements}

%
%

\bibliographystyle{aa}
\bibliography{./Refs/journals,./Refs/psrrefs,./Refs/modrefs,./Refs/crossrefs} 



\end{document}